\documentclass[12pt]{article}

\usepackage{amssymb}
\usepackage[fleqn]{amsmath}
\usepackage{amsthm}
\usepackage{geometry}
\usepackage{fleqn}
\usepackage{xspace}
\usepackage{color}
\usepackage{xcolor}
\usepackage{graphicx}
\usepackage{mathrsfs}
\usepackage[colorlinks=true,linkcolor=blue,citecolor=blue,urlcolor=blue]{hyperref}
\usepackage{dutchcal}

\theoremstyle{definition}

\newtheorem{definition}{Definition}[section]
\newtheorem{theorem}[definition]{Theorem}

\newtheorem{lemma}[definition]{Lemma}
\newtheorem{proposition}[definition]{Proposition}
\newtheorem{example}[definition]{Example}

\newcommand{\done}{\hfill\ensuremath{\Box}}
\newcommand{\Tuple}[1]{\ensuremath{\big\langle #1\big\rangle}\xspace}
\newcommand{\tuple}[1]{\ensuremath{\langle #1\rangle}\xspace}
\renewcommand{\implies}{\rightarrow}

\newcommand{\asp}{\textsc{Asp}\xspace}
\newcommand{\sql}{\textsc{Sql}\xspace}
\newcommand{\golog}{\textsc{Golog}\xspace}
\newcommand{\krr}{\textsc{Krr}\xspace}
\newcommand{\dls}{\textsc{Dls}\xspace}
\newcommand{\nptime}{\textsc{Np}\xspace}
\newcommand{\conptime}{\textsc{co-Np}\xspace}
\newcommand{\ptime}{\textsc{P}\xspace}
\newcommand{\logspace}{\textsc{LogSpace}\xspace}

\newcommand{\cala}{\ensuremath{\mathcal{A}}\xspace}

\newcommand{\calc}{\ensuremath{\mathcal{C}}\xspace}

\newcommand{\calr}{\ensuremath{\mathcal{R}}\xspace}
\newcommand{\cals}{\ensuremath{\mathcal{S}}\xspace}
\newcommand{\calt}{\ensuremath{\mathcal{T}}\xspace}
\newcommand{\calv}{\ensuremath{\mathcal{V}}\xspace}
\newcommand{\calw}{\ensuremath{\mathcal{W}}\xspace}

\newcommand{\defeq}{\stackrel{\mathrm{def}}{=}}
\newcommand{\bydefequiv}{\stackrel{\mathrm{by\, def}}{\equiv}}
\newcommand{\defequiv}{\stackrel{\mathrm{def}}{\equiv}}

\newcommand{\lneg}{\neg}

\newcommand{\true}{\ensuremath{\mathbb{T}}\xspace}
\newcommand{\false}{\ensuremath{\mathbb{F}}\xspace}

\newcommand{\abstr}{\mbox{$\alpha$-ab}\-strac\-tion\xspace}
\newcommand{\Abstr}{\mbox{$\alpha$-Abstraction}\xspace}

\newcommand{\snc}[3]{\ensuremath{\mathit{snc}^{#3}\big(#1\,|\,#2\big)}\xspace}
\newcommand{\wsc}[3]{\ensuremath{\mathit{wsc}^{#3}\big(#1\,|\,#2\big)}\xspace}

\newcommand{\ef}{\ensuremath{\mathit{ef}}\xspace}
\newcommand{\bc}{\ensuremath{\mathit{bc}}\xspace}
\newcommand{\spp}{\ensuremath{\mathit{sp}}\xspace}
\newcommand{\ecs}{\ensuremath{\mathit{ecs}}\xspace}

\newcommand{\mood}{\mathit{mood}}
\newcommand{\play}{\mathit{play}}
\newcommand{\enjoy}{\mathit{enjoy}}
\newcommand{\game}{\mathit{game}}

\newcommand{\tim}{\mathit{tim}}
\newcommand{\rep}{\mathit{rep}}
\newcommand{\lgh}{\mathit{lgh}}
\newcommand{\slm}{\mathit{slm}}
\newcommand{\mnt}{\mathit{mnt}}
\newcommand{\vel}{\mathit{vel}}

\begin{document}

 \title{ Bridge and Bound:\\
 	 A Logic-Based Framework for  Abstracting\\ (Extended Report)}
 
\date{}

 \author{Andrzej Sza{\l}as\\
 \normalsize  Institute of Informatics, University of Warsaw\\ 
 \normalsize  Banacha 2, 02-097 Warsaw, Poland
}

 \maketitle
 
 \begin{abstract}
	At its core, abstraction is the process of generalizing from specific instances to broader concepts or models, with the primary objective of reducing complexity while preserving properties essential to the intended purpose. It is a~fundamental, often implicit, principle that structures the understanding, communication, and development of both scientific knowledge and everyday beliefs. Studies on abstraction have  evolved from its origins in Ancient Greek philosophy through methodological approaches in psychological and philosophical theories to modern computational frameworks. 
	
	This paper presents a novel logic-based framework for modeling abstraction processes in which all components are expressed within logic. The framework extends beyond the traditional focus on the entailment of necessary conditions by making sufficient conditions first-class citizens as well.  We define approximate abstractions, study their tightest and exact forms, and extend the approach to layered abstractions, enabling hierarchical simplification of complex systems and models. The computational complexity of the related reasoning tasks is also discussed. 
	
	For clarity, our framework is developed within classical logic, chosen for its simplicity, expressiveness, and computational friendliness.
\end{abstract}

\subsubsection*{Keywords}
	abstracting; classical propositional logic; classical first-order logic; classical second-order logic; approximate theory;  strongest necessary condition; weakest sufficient condition; quantifier elimination

\vfil
\eject

\section{Introduction}\label{sec:intro}

\subsection{Abstracting}

Abstraction is a basic and often implicit concept that influences how we understand, communicate, and develop both scientific knowledge and everyday beliefs. From commonsense efforts to categorize the real world to the formation of advanced scientific theories and models, abstraction is omnipresent in human inquiry. Scientific theories are highly abstract frameworks unifying observations and laws under a~set of general principles, allowing for reasoning far more efficient than deliberating over minor details. 
The concept of abstraction has evolved significantly from Ancient Greek philosophy to  modern theoretical frameworks~\cite{aristotles-abstr,abstractionism,abstr-neo-fregean}. In computer science, abstraction often functions as a primary design and research method in software engineering, particularly in object-orientation, modular programming methodologies, abstract data types, software verification or model checking~\cite{ClarkeGL94,cousot-book,Turner21}. However, in order to maintain a~manageable and coherent scope, we will limit the focus of this paper to modeling the process of abstraction, primarily inspired by subdisciplines of \krr that emphasize knowledge/belief representation and reasoning, ontologies, data integration, querying and rule-based languages~\cite{Cima-lics,CimaPL23,GiunchigliaW92,Luo0LL20-sitcalc,LutzS23,NayakL95,asp-abstraction}, just to mention a~few of related approaches.\footnote{Given the vast volume of existing literature on abstracting, even within the paper's narrowed scope, providing a comprehensive list of references is unfeasible. Thus, we cite selected works, and refer the readers to bibliographies included there. However, all foundational works upon which our model is constructed are explicitly cited whenever used.} 
It is, in particular, used to mitigate the computational complexity inherent in tasks like automated reasoning, planning, etc. Frameworks grounded in both classical and non-classical logics play a~prominent role in these areas, particularly in tasks such as modeling real-world environments, reasoning, specification and verification of complex systems, and querying data/belief/knowledge bases.   

At its core, abstraction involves the process of generalizing from detailed instances to form broader concepts or models with the goal to reduce the complexity of further reasoning, communicating and developing new knowledge layers. That is, abstraction can be understood as a~process of transforming a \emph{source} representation (theory) specifying a problem into an \emph{abstract} representation (theory) that throws away certain details while retaining specific desirable properties.\footnote{In the literature, source theories/representations are also called  \emph{ground}, \emph{base} or \emph{concrete}. We use the term \emph{source} representation (theory) to emphasize its role as a starting point for abstracting, regardless of whether it originates from raw, concrete reality or from a previously established, lower-level abstraction.}
A key component of an abstraction process is an abstract language, the choice of which is closely guided by the nature of queries the resulting abstract system or model is intended to address, aiming to capture essential features and behaviors relevant to the research question. For the clarity of presentation, we use classical logic, a relatively simple yet expressive, powerful, computationally friendly and illustrative formalism. In this paper, any classical propositional/first-order theory is expressed by a finite set of formulas, understood as a~single formula formed by their conjunction. Depending on whether we work within a~propositional or first-order setting, the vocabulary of a~theory consists of either propositional variables or relation symbols, respectively. We assume that the \emph{details to be abstracted away} constitute a subset of the considered vocabularies.  While we focus on classical logic, the proposed methodology can be adapted to non-classical logics and logic-based reasoning techniques and engines.

Since classical logic serves as this paper's formalism, the languages within our framework are defined by establishing their vocabularies. Consequently, we assume that the abstract language (vocabulary) is specified during the abstraction process, and is determined by the theorist.\footnote{Across the relevant disciplines, there is no standardized title for the individual performing abstraction. In this paper, we adopt the term \emph{theorist} as referring to this role.} Given the source vocabulary $\calv_S$, an abstract vocabulary, $\calv_A$, is established by identifying specific elements within $\calv_S$ for generalization or omission at a higher (more abstract) level of representation, $\calr_A$. The vocabulary $\calv_A$ introduces new symbols, not present in $\calv_S$, that encapsulate generalized concepts derived from the source.

Given the source theory $\calr_S$ and an abstract vocabulary $\calv_A$, we conceptualize abstraction, as a~process comprising the following major steps:
\begin{itemize}
\item establishing a bridging layer: an intermediate formal layer, referred to as the \emph{bridging theory}  $\calt_B$, is constructed by the theorist to formally capture the relationships between the source theory $\calr_S$  and abstract (approximate) theory~$\calr_A$; 
\item deriving the target (abstract) representation: an approximate theory $\calr_A$ is derived as an abstract representation to ensure the preservation of selected desired meta-properties;\footnote{According to tradition, we distinguish between \emph{properties} expressible by formulas of a given logic, and \emph{meta-properties} about the logic, which are typically not expressible by its formulas.} in the paper these meta-properties are related to the traditional entailment of necessary conditions as well as the entailment by sufficient conditions.\footnote{By selecting the mentioned meta-properties, in addition to standard logical reasoning involving necessary conditions, we address reasoning with sufficient conditions useful, among others, in abduction or explanation~\cite{dlssnc,Lin,Lin01}.} 
\end{itemize}
The abstract representation $\calr_A$  will be obtained by projecting the source representation $\calr_S$ onto the abstract vocabulary $\calv_A$ under the bridging theory $\calt_B$. As we will show, such projections are typically approximate rather then exact. The representation $\calr_A$ will then be obtained by \emph{bounding} which delivers two bounds being theories: the first preserving the entailment of necessary conditions, and the second one preserving the entailment by sufficient ones.

The techniques formalizing the above steps are central to our branch-and-bound framework and are briefly discussed below.

\subsection{Bridging}

As brought up above, by \emph{bridging} we understand a construction of  a more or less explicit formalism that connects lower and higher levels of abstraction. While many approaches provide forms of bridging at a meta-level, i.e., by employing a language external to both the source and abstract languages and tailored to ensure crucial meta-properties, our approach is unique in that it assumes that bridging is specified within the object formalism itself.

In existing logic-based models, abstraction process is typically encapsulated within a mapping between the source and the abstract system. A prominent example is the famous abstraction principle applied in mathematics.\footnote{Though general Frege abstraction principle (Basic Law V) appeared inconsistent, there are its consistent repairs~\cite{abstractionism,abstr-neo-fregean} used in the foundations of mathematics.} The idea is to identify equivalent elements, so bridging is achieved through a mapping that assigns each object to a common representative, such that all objects equivalent under the relation are mapped to the same representative of the equivalence class. For instance, one may identify parallel straight lines to obtain a more abstract concept of direction. 
Mapping-based approaches to abstraction are also clearly present in the \krr area. To illustrate the broad spectrum of such approaches, consider some well-known works. In~\cite{GiunchigliaW92}, a general logic-based theory of abstraction is proposed, encompassing many different approaches. It is defined as a pair of formal deductive first-order systems: the source system and the abstract target system,	together with an effective mapping from the source language to the target language. This mapping serves to bridge the two systems. In~\cite{Luo0LL20-sitcalc}, bridging is defined through a refinement mapping that maps source to abstract fluents, and action functions to \golog programs. Then sound and complete abstractions, related to forgetting~\cite{Lin94forgetit}, are investigated. In~\cite{SaribaturEiter21}, omission-based abstraction is studied in the context of Answer Set Programming (\asp). The bridging is defined by requiring the existence of an abstraction mapping between ground atoms (not containing first-order variables)  of source and abstract \asp programs that preserves answer sets.\footnote{\asp programs are finite sets of rules allowing, among others, for non-monotonic reasoning~\cite{BrewkaET11,gekakasc12a,GelKah:2014}.} Over-approximation is achieved so that every necessary condition of an abstract \asp program is also a~necessary condition of the corresponding source program. In~\cite{NayakL95}, bridging is given by mappings between models of the source and abstract theory so that  a consequence of the abstract theory (necessary condition) has its counterpart among the necessary conditions of the source theory. In~\cite{cousot-book} and  papers referenced there, in particular~\cite{CousotC79}, bridging is mainly achieved through Galois connections or, alternatively, topological closures that formalize the correspondence between concrete and abstract semantics of programs. Naturally, these are only illustrative examples; however, they demonstrate the concept of bridging within selected formalisms frequently considered in the literature.

While sharing many underlying intuitions with the cited works, our approach generalizes mappings between languages to bridging theories, which do not necessarily determine specific mappings. We believe that this approach models the process of abstraction more directly and naturally.

\subsection{Bounding}

In the process of modeling real-world phenomena, idealizations and approximations are typically unavoidable~\cite{sep-models-science}. For example, when considering a~traffic situation and having access to parameters such as road slipperiness, air humidity and temperature, speed limits, the actual speed of the vehicle, and so on, one often prefers to operate with more abstract concepts, such as an ``(un)safe situation''. For such complex concepts, it is possible to specify sufficient conditions that imply them, as well as necessary conditions that they imply. However, in many cases it is practically impossible to enumerate all such conditions and/or provide a precise definition of these situations. Therefore, in applications, such concepts are approximated by selected sufficient and necessary conditions, which are then maintained and extended while used in reasoning processes. Consequently, it should not be surprising that the specification of the bridging relationship between source and abstract theories does not uniquely determine the latter, making approximations a common and significant feature when bridging is concerned.\footnote{Whenever we use the notion \emph{uniquely determined}, it is to be understood as \emph{uniquely determined up to equivalence}. More precisely, we implicitly work in the Lindenbaum–Tarski algebra, whose elements are equivalence classes of formulas, rather than the formulas themselves.} The question is then how the related approximations are to be understood? In the framework developed in this paper, we concentrate on approximations that address two fundamental aspects: (i) reasoning from an abstract representation that preserves as much as possible of the reasoning from the source representation, i.e., preserving as much as possible of the necessary conditions; and (ii) identifying sufficient conditions for the abstract representation that retain as much as possible of the sufficient conditions for the source representation. Abstractions will then be approximated by providing two bounds: an \emph{upper bound} related to (i) and a \emph{lower bound} related to (ii). In essence, our approach to modeling the process of abstraction can be summarized as \emph{bridge and bound}. That is, given the source representation and an abstract vocabulary, the theorist provides a theory bridging source and abstract entities, and our bridge-and-bound framework offers tools to express bounds that approximate source formalism-based reasoning within the abstract formalism. In particular, it allows one to automatically find the best (\emph{tightest}) such bounds.

\subsection{Contents and Structure of the Paper}

The original contents of the paper includes:
\begin{itemize}
\item a novel bridge-and-bound framework for abstracting, focusing not only on traditional entailment of necessary conditions, but also taking into account the entailment by sufficient ones;
\item approximate abstractions (\abstr{s}) given by upper and lower bounds that correspond to necessary and sufficient conditions as well as tightest and exact such approximate abstractions;
\item a bridge-and-bound approach to layered abstractions enabling the simplification of complex systems by organizing them into hierarchical layers;
\item a discussion of computational techniques for determining the best bounds for the target tightest abstractions.
\end{itemize}
While abstraction-based approximations have been considered previously,e.g., in~\cite{cousot-book,CousotC79,SaribaturEiter21,asp-abstraction}, existing approaches focus primarily on the inference of necessary conditions, that is, upper bounds. In particular, \cite{SaribaturEiter21,asp-abstraction} discusses over-approximation obtained by admitting abstract answer sets that are not answer sets of the original program, thereby ensuring soundness of reasoning. In~\cite{cousot-book,CousotC79}, approximation is also considered, but at the level of fixpoints used to formalize program semantics. To the best of our knowledge, no prior work investigates reasoning based on sufficient conditions (lower bounds), nor addresses the \abstr, and in particular \emph{best} (i.e., \emph{tightest}) bounds for such abstractions.

It is also worth emphasizing that our bridge-and-bound framework does not rely on mappings between source and abstract systems; it only requires bridging theories that establish, possibly implicitly, the connection between source and abstract concepts. Moreover, all components of the branch-and-bound framework are fully expressed in logic, whereas other approaches typically rely on theories external to the base logical language. This provides a unique perspective and a uniform framework, allowing for reasoning both with and about abstraction in a uniform manner.

The rest of the paper is organized as follows. Section~\ref{sec:logic} reviews the necessary logical background relevant to the paper.  Section \ref{sec:abstract} introduces the bridge-and-bound framework, defines the notions of \abstr{s} along with their tightest and exact forms, and discusses layered abstractions while analyzing their properties. In particular, it formally characterizes tightest abstractions using the strongest necessary and weakest sufficient conditions, provides a sufficient condition for \abstr to be exact, and demonstrates a form of compositionality for layered \abstr. Section~\ref{sec:fo} extends this framework to the first-order setting. In Section~\ref{sec:compaspects} we discuss the corresponding computational techniques and complexity of the approach. Finally, Section~\ref{sec:concl} concludes the paper with final remarks and outlines potential directions for future research.

\section{Logical Background}\label{sec:logic}

\subsection{Classical Propositional Logic}

For the sake of simplicity, the major ideas will be presented by focusing on classical propositional logic with truth values \true (true) and \false (false), an enumerable set of propositional variables, $\calv^0$, and standard connectives of negation, conjunction, disjunction, implication and equivalence, $\lneg,\land,\lor,\implies,\equiv$. A \emph{literal} is a~propositional variable or its negation, the former called a \emph{positive} and the latter  a~\emph{negative literal}.

A \emph{theory} is a finite set of propositional formulas.\footnote{Since we do not assume that theories are deductively closed, such finite sets of formulas are frequently called \emph{belief bases}.\label{footnote:dedclosed}} As usual, such propositional theories are identified with conjunctions of their formulas, so we use the terms ``formula'' and ``theory'' interchangeably. Traditionally, for the boundary case when theory $\calt=\emptyset$, we assume that \calt is equivalent to \true. 

By a \emph{vocabulary} (a \emph{signature}) of a~theory \calt we mean the set of all propositional variables occurring in \calt. Since theories are finite, the considered vocabularies are finite, too. Writing $\calt(\calv)$ or $\calt(\calv,\calw)$ we indicate that the vocabulary of theory \calt is (exactly) \calv or, respectively, $\calv\cup\calw$, where in the latter case we always assume that $\calv\cap\calw=\emptyset$. 

By a~\emph{world} over vocabulary \calv we mean any assignment $w$ of truth values to variables in \calv, $w:\calv\longrightarrow \{\false,\true\}$. The assignment $w$ can be extended to all formulas inductively in a standard way, where $A, B$ are propositional formulas:
\begin{itemize}
\item $w(\lneg A)\defeq \true$ if $w(A)=\false$; $w(\lneg A)\defeq \false$ if $w(A)=\true$;
\item assuming that we have the ordering $\leq_0$ on truth values, where $\false<_0\true$, we define:\footnote{Given the ordering $\leq_0$, the relationships are defined in a standard way, $<_0$ as the relation of being strictly smaller, $\min_{\leq_0}$, $\max_{\leq_0}$ respectively as the minimum and maximum with respect to $\leq_0$, and $=_0$ as the equality with respect to  $\leq_0$, i.e., being both $\leq_0$, and $\geq_0$.}\\ $w(A\land B)\defeq \min_{\leq_0}\{w(A),w(B)\}$, $w(A\lor B)\defeq \max_{\leq_0}\{w(A),w(B)\}$, \\
$w(A\implies B)\defeq \true$ iff $w(A)\leq_0 w(B)$, $w(A\equiv B)\defeq \true$ iff $w(A)=_0 w(B)$.
\end{itemize}

For a world $w$ over $\calv$, and a theory $\calt$ over vocabulary included in \calv, we say that $w$ is a \emph{model} of $\calt$, denoted by $w\models \calt$, when $w(\calt)=\true$. Theory $\calt'(\calv')$ \emph{entails} theory $\calt''(\calv'')$, denoted by  $\calt'\models\calt''$, iff for every world over vocabulary $\calv'\cup\calv''$ we have that $w\models \calt'$ implies $w\models\calt''$. We abbreviate $\emptyset\models\calt$ to $\models\calt$. Since the empty theory is identified with \true, this means that \calt, being true in all models, is a~\emph{tautology}. 

\subsection{Strongest Necessary and Weakest Sufficient Conditions}

By \emph{a necessary condition over vocabulary $\calw$ of a formula $A(\calv,\calw)$  under theory $\calt(\calv,\calw)$} we shall understand any formula $B(\calw)$ such that: 
\[\calt(\calv,\calw)\models  A(\calv,\calw)\implies B(\calw).\] 
Such a formula $B(\calw)$ is the  \emph{strongest necessary condition of $A(\calv,\calw)$  over vocabulary \calw under theory $\calt(\calv,\calw)$}, denoted by $\snc{A(\calv,\calw)}{\calt(\calv,\calw)}{\calw}$, if additionally,
for any necessary condition $C(\calw)$ of $A(\calv,\calw)$ on $\calw$ under $\calt(\calv,\calw)$, we have that  $\calt(\calv,\calw)\models  B(\calw)\implies C(\calw).$

By \emph{a sufficient condition over vocabulary $\calw$ of a formula $A(\calv,\calw)$  under theory $\calt(\calv,\calw)$}  we shall understand any formula
$B(\calw)$ such that:
\[\calt(\calv,\calw)\models B(\calw)\implies  A(\calv,\calw).\] 
Such a formula $B(\calw)$ is the \emph{weakest sufficient condition  of $A(\calv,\calw)$ over vocabulary \calw under theory  $\calt(\calv,\calw)$}, denoted by $\wsc{A(\calv,\calw)}{\calt(\calv,\calw)}{ \calw}$, if additionally, for any sufficient condition $C(\calw)$ of $A(\calv,\calw)$ on $ \calw$ under $\calt(\calv,\calw)$, we have that  $\calt(\calv,\calw)\models C(\calw)\implies B(\calw)$.

Weakest sufficient and strongest necessary conditions for classical propositional logic, as defined above, have been introduced in~\cite{Lin} (for an extended version of~\cite{Lin}, see~\cite{Lin01}) and further extended to first-order case in~\cite{dlssnc}. Let for a propositional formula $B$, $A(p=B)$ denote the formula obtained from formula $A$ by substituting all occurrences of propositional variable $p$ in $A$ by $B$. For characterizing $snc()$ and $wsc()$, we use second-order quantifiers $\exists p, \forall p$ ($p\in\calv^0$),  defined by:
\begin{align}
&\exists p\big(A(p)\big)\defequiv A(p=\false)\lor A(p=\true);\label{eq:exists}\\
&\forall p\big(A(p)\big)\defequiv A(p=\false)\land A(p=\true),\label{eq:forall}
\end{align} 
The following second-order characterization of $snc()$ and $wsc()$ has been provided in~\cite{dlssnc}, where for $\calv=\{p_1,\ldots,p_k\}$, notation $\exists\calv$ and $\forall\calv$ stands for $\exists p_1\ldots\exists p_k$ and $\forall p_1\ldots\forall p_k$, respectively.

\begin{lemma}[Lemma 3.1 in~\cite{dlssnc}]\label{lemma:wsc-snc-so}
For any $ A(\calv,\calw)$, $ \calw$ and $\calt(\calv,\calw)$:
\begin{align}
	&\snc{A(\calv,\calw)}{\calt(\calv,\calw)}{ \calw}\equiv \exists  \calv\big(\calt(\calv,\calw)\land  A(\calv,\calw)\big);\\
	&\wsc{A(\calv,\calw)}{\calt(\calv,\calw)}{ \calw} \equiv \forall  \calv\big(\calt(\calv,\calw)\implies  A(\calv,\calw)\big). 
	\label{eq:so-wsc}
\end{align}
\mbox{}\\[-2.4em]

\done
\end{lemma}

\noindent Obviously, $snc()$ and $wsc()$ are dual to each other. That is, 
\[
\begin{array}{l}
	\snc{A(\calv,\calw)}{\calt(\calv,\calw)}{ \calw}\equiv\lneg\wsc{\lneg A(\calv,\calw)}{\calt(\calv,\calw)}{ \calw};\\
	\wsc{A(\calv,\calw)}{\calt(\calv,\calw)}{ \calw}\equiv\lneg\snc{\lneg A(\calv,\calw)}{\calt(\calv,\calw)}{ \calw}.	
\end{array}
\]

\noindent We consider both $snc()$ and $wsc()$ for two main reasons. First, their use helps keep the presentation clear and much easier to follow. Second, the availability of both operators allows the framework to be applied to positive (negation-free) fragments of classical logic, and also can serve as a natural starting point for adapting the framework to non-classical logics including those not admitting negation.

\subsection{Approximate Theories and Bounds}

\begin{figure}[t] 
\centering 
\includegraphics[width=0.62\textwidth]{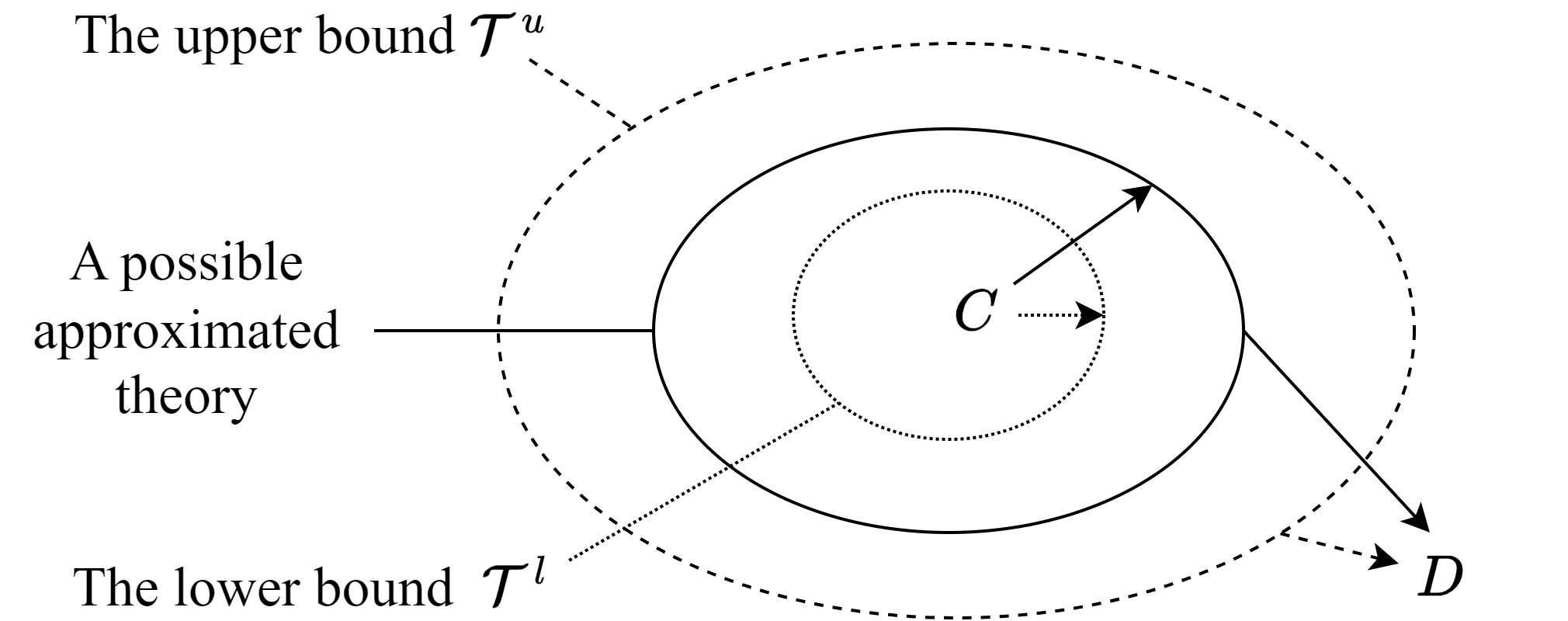} 
\caption{Approximate theory \tuple{\calt^l,\calt^u}: any sufficient condition $C$ for $\calt^l$ and any necessary condition $D$ of $\calt^u$ is also a sufficient (respectively, necessary) condition of any theory located between bounds $\calt^l$ and $\calt^u$  (with respect to entailment). Arrows indicate implications.}
\label{fig:approx}
\end{figure}

In the rest of the paper rather than to theories, we shall frequently refer to \emph{approximate theories}, abbreviated to \emph{approximations}. Assuming that a background theory $\calt$ holds, \emph{approximations under \calt over vocabulary} \calw, are defined as pairs of theories $\cala\defeq\tuple{\calt^l,\calt^u}$  over vocabulary \calw, where $\calt^l$ is called the \emph{lower bound} of $\cala$,  $\calt^u$~is called its \emph{upper bound}, and we require that  $\calt\models\calt^l\implies\calt^u$.  The role of $\calt^l$ is to encapsulate (though possibly not all) sufficient conditions of the approximate theory, whereas $\calt^u$ encapsulates (again, possibly not all) its necessary conditions (see Figure~\ref{fig:approx}). To maintain uniformity in presentation, we shall understand any theory $\calt$ as the approximation $\tuple{\calt,\calt}$ under the empty theory (being \true). The approximation $\cala=\tuple{\calt^l,\calt^u}$  is called \emph{exact } under \calt  iff $\calt\models \calt^l\equiv \calt^u$. 

For any theories \calt and  $\calt'$, the approximation $\cala=\tuple{\calt^l,\calt^u}$ is called the \emph{tightest  approximation of $\calt'$  over vocabulary} \calw under \calt  iff every sufficient condition for $\calt'$ on \calw under \calt is a sufficient condition for $\calt^l$, and every necessary condition  of $\calt'$ on  \calw under \calt is a~necessary condition of $\calt^u$. 

By an obvious application of transitivity of implication, sufficient conditions always imply the necessary ones.  Using this fact and the definition of $snc()$ and $wsc()$, we have the following lemma.

\begin{lemma}\label{lemma:sncwsctapprox}
Let $\calt'(\calv,\calw), \calt(\calv,\calw)$ be arbitrary propositional theories, and let  $\cala\defeq\Tuple{\wsc{\calt'(\calv,\calw)}{\calt(\calv,\calw)}{ \calw},\snc{\calt'(\calv,\calw)}{\calt(\calv,\calw)}{ \calw}}$. Then: 
\begin{itemize}
	\item \cala is an approximation of $\calt'(\calv,\calw)$ under  $\calt(\calv,\calw)$ over vocabulary \calw;
	\item \cala is the tightest (with respect to entailment) approximation of $\calt'(\calv,\calw)$ under  $\calt(\calv,\calw)$ over vocabulary \calw.	\done
\end{itemize}
\end{lemma} 

\subsection{Remarks}

We consider approximations due to factors both internal and external to the scope of this paper. The internal motivation arises from the observation that abstractions may be approximations rather than single (exact) theories. 

The external motivation comes from the methods of logic-based modeling of reality, such as used in \krr including rule-based languages and other reasoning engines, where it often appears impractical or impossible to specify a~single, ``ideal'' theory. Instead, one typically constructs  approximations of the intended theory. This phenomenon is particularly evident in rule-based systems, where,  introducing a~new concept (say, $c$), it is customary to define both sufficient conditions for $c$ and for its complement:\footnote{For simplicity, we use implications rather than rules of rule-based languages.}
\begin{align}
\!\!\!\!(c_1\land\ldots\land c_m)\implies c\;\;\, & \mbox{\;\;\;\;with $c_1\land\ldots\land c_m$ being a sufficient condition for $c$;}\label{eq:ortho-th-1}\\
\!\!\!\!(d_1\land\ldots\land d_n)\implies \lneg c & \mbox{\;\;\;\;equivalent to\ } c\implies \lneg (d_1\land\ldots\land d_n), \mbox{ therefore}\label{eq:ortho-th-2}\\
\!\!\!\!&\mbox{\;\;\;\;with $\lneg (d_1\land\ldots\land d_n)$ being a necessary condition for $c$.}\nonumber
\end{align}
In \eqref{eq:ortho-th-1}--\eqref{eq:ortho-th-2}, propositional variable $c$ can represent a given state, such as the ``safe situation'' mentioned in Section~\ref{sec:intro}. Conjunctions of variables $c_1,\ldots, c_m$ may represent sufficient conditions for a safe situation $c$ (e.g., ``dry road'', ``low speed'', etc.), while conjunctions of $d_1,\ldots, d_n$ -- sufficient conditions for an unsafe situation $\lneg c$ (e.g., ``slippery road'', ``high speed'').

Additionally, one may specify further implications involving an arbitrary  literal, say $\ell$, as a potential conclusion: 
\begin{align}
\!\!\!\!(e_1\land \ldots\land c\land \ldots\land e_k)\implies \ell\;\; & \mbox{\;\; equivalent to\ }
c\implies (\ell\lor \lneg e_1\lor \ldots\lor \ldots\lor \lneg e_k);\\
\!\!\!\!(f_1\land \ldots\land\lneg c\land\ldots\land f_l)\implies \ell & \mbox{\;\; equivalent to\ } 
(f_1\land \ldots\land\lneg \ell\land\ldots\land f_l)\implies c.
\end{align}
In essence, each of these implications contributes, in some form, to the specification of either necessary or sufficient conditions for $c$. Of course, this informal exposition is only intended to highlight how such theories are constructed in the broader practice of logic programming paradigms.\footnote{Though equivalence-based definitions are often implicitly present in logic programming and rule-based query languages, this is achieved by applying the simplifying assumption that the specified rules provide \emph{all} sufficient conditions~\cite{AHV,log-programming}. While compatible with the least/minimal model semantics, in many cases it might appear not fully adequate~\cite{PM2023}.}  Nevertheless, the underlying intuition generalizes to many other knowledge representation formalisms.

Soft computing, particularly in the context of approximate reasoning, also supports this perspective. For example, orthopairs~\cite{Ciucci11,ciucci16} encompassing many approximate reasoning techniques, are closely related to approximate theories as understood in our paper. An \emph{orthopair} is a pair of disjoint subsets of a given universe $\tuple{P,N}$
with $P$ gathering positive and $N$ -- negative examples, like positive and negative literals, accepted or rejected examples or objects in machine learning, etc. This view is compatible with intuitions behind~\eqref{eq:ortho-th-1} and \eqref{eq:ortho-th-2}. Also, given an orthopair \tuple{P,N}, with $P,N$ consisting of literals, one can construct the related approximation, \tuple{P,N^c}, where $X^c$ denotes the complement of a~set $X$. Orthopairs can model many other approaches, including perhaps the most prominent among them -- rough sets, introduced in~\cite{PZ81} and surveyed in many other books including, e.g., ~\cite{dlss2006a,Pa91,polkowski,rough-book}. In rough sets lower and upper approximations are constructed on the basis of equivalence relation modeling indistinguishability of domain objects.\footnote{In our paper we approximate theories rather than sets. Also we deal with a rather weak requirement that the lower bound is included in the upper bound, what would rather correspond to the seriality of the underlying similarity relation rather than much stronger equivalence.} 

Another important area in which theories are approximated in a manner similar to that used in our paper is knowledge compilation~\cite{CadoliD97,DarwicheM24,kautzselmanK96}, where one approximates theories by less complex, typically tractable ones that bound the original theory from below and above. 

Summing up, the concept of approximate theories is motivated both internally, by the tendency of abstractions to naturally take this form in the presented framework, and externally, by practices in logic-based modeling where specifying a single ideal theory is often impractical or even impossible. Instead, partial or approximate theories are constructed using sufficient and necessary conditions. This idea also aligns with approaches in soft computing, such as orthopairs and rough sets. Additionally, approximations are conceptually related to knowledge compilation, where complex theories are approximated by simpler and less computationally demanding  bounds.

\section{A Framework for Abstracting}\label{sec:abstract}

\subsection{Approximate Abstractions}

The following definition formalizes abstractions as pairs consisting of a lower and an upper bound, as already announced in Sections~\ref{sec:intro} and~\ref{sec:logic}.

\begin{definition}[Approximate abstraction]\label{def:a-abstr}
By an \emph{approximate abstraction} (further referred to as \emph{\abstr}) \emph{from the source theory $\calt_S$ with respect to the bridging theory $\calt_B$ over vocabulary $\calv_A$} we mean any pair $\tuple{\calt^l, \calt^u}$, where $\calt^l$ and $\calt^u$, respectively called its \emph{lower} and \emph{upper bound}, are theories over $\calv_A$ such that:
\begin{itemize}
	\item $\calt^l$ preserves the entailment by sufficient conditions over $\calv_A$ under theory $\calt_B$; that is, for any formula $C$ on the vocabulary $\calv_A$,
	$\calt_B\models C\implies \calt^l$ implies $\calt_B\models C\implies \calt_S$;
	\item $\calt^u$ preserves the entailment of necessary conditions over $\calv_A$ under theory $\calt_B$; that is, for any formula $D$ on the vocabulary $\calv_A$, $\calt_B\models \calt^u\implies D$ implies $\calt_B\models \calt_S\implies D$.
	\done
\end{itemize}
\end{definition}

It is important to notice that \abstr{}s do not have to be uniquely determined. For example, $\tuple{\false,\true}$ is always a (trivial) \abstr.

\begin{figure}[t] 
\centering 
\includegraphics[width=0.99\textwidth]{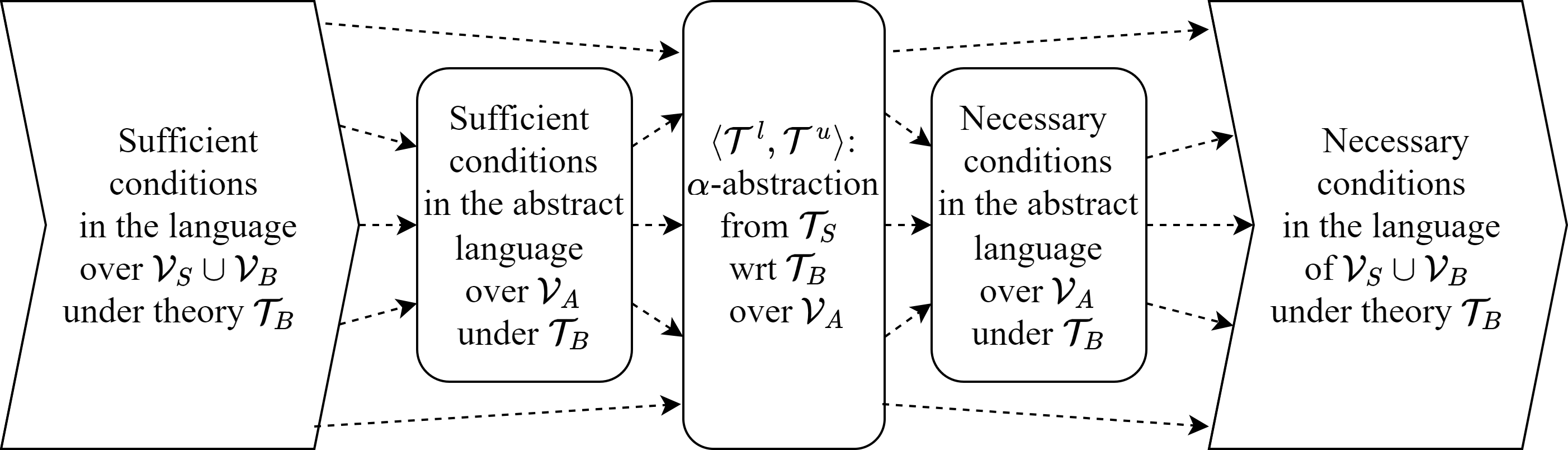} 
\caption{The role of theories in abstracting: $\calt_S$ is a source theory, $\calt_B$ is a bridging theory over vocabularies $\calv_S$ and $\calv_B$, respectively, $\calv_A$ is an abstract vocabulary, and $\calt^l,\calt^u$ are a lower and an upper bound of an~\abstr. Dashed arrows represent implications. }
\label{fig:abstr}
\end{figure}

The relationships among theories participating in abstracting, as formalized in Definition~\ref{def:a-abstr},  are illustrated in Figure~\ref{fig:abstr}.

The following proposition indicates that \abstr{}s are approximate theories.

\begin{proposition}\label{prop:abstr-approx-th}
Let $\cala=\tuple{\calt^l, \calt^u}$ be an \abstr from any source theory with respect to a~bridging theory $\calt_B$ and an arbitrary vocabulary. Then \cala is an approximate theory under $\calt_B$.\done
\end{proposition}

The following example  briefly illustrates our approach. 

\begin{example}\rm\label{ex:game-mood}
Consider a simple scenario in which a game player is in a mood for a~new game ($\mood$), and this particular mood makes the player  interested in games $g_1$ and $g_2$, planning to purchase one or both. Regardless of which game is purchased, the player will play ($\play$) and expects to enjoy  the game ($\enjoy$). The player's source theory, $\calt_p$, consists of the following formulas:
\begin{align}
	& \mood\implies (g_1\lor g_2);\label{eq:player1}\\
	& (g_1\lor g_2)\implies (\play\land \enjoy).\label{eq:player2}
\end{align}
Suppose the player wants to simplify and slightly generalize the reasoning by abstracting away from the specific games $g_1$, $g_2$, and using a general term $\game$ instead. The player then considers the following bridging theory, $\calt_g$: 
\begin{equation}\label{eq:playerbridge}
	(g_1\lor g_2) \implies \game.
\end{equation}
It is important to note that Equation~\eqref{eq:playerbridge} offers an implicit, potentially ambiguous formulation of the relationship between the source theory $\calt_p$ and its \abstr{}s. That is, an \abstr does not have to be uniquely determined by \eqref{eq:player1}, \eqref{eq:player2} under \eqref{eq:playerbridge}. For example, one could consider \abstr{}s: 
\begin{itemize}
	\item $\cala_1=\tuple{\lneg \mood\land\lneg\game, \mood\implies \game}$; indeed:
	\begin{itemize}
		\item the formula $\lnot \mood \land \lnot \game$ is a lower bound for $\calt_p$, being its sufficient condition under~\eqref{eq:playerbridge}, since $\lnot \mood$ makes~\eqref{eq:player1} true, and $\lnot \game$ implies, by~\eqref{eq:playerbridge},  $\lnot(g_1 \lor g_2)$, which in turn makes~\eqref{eq:player2} true;
		\item the formula $\mood \implies \game$ is an upper bound for $\calt_p$, being its necessary condition, what  is evident from the transitivity of implication applied to~\eqref{eq:player1} and~\eqref{eq:playerbridge};
	\end{itemize}
	\item $\cala_2=\tuple{\lneg\mood\land(\play\land\enjoy),\mood\implies(\play\land\enjoy)}$; indeed:
	\begin{itemize}
		\item the formula $\lnot \mood \land(\play\land\enjoy)$ is a lower bound for $\calt_p$, as it is its sufficient condition under~\eqref{eq:playerbridge}: as before, $\lnot \mood$ makes~\eqref{eq:player1} true, and $\play\land\enjoy$ makes true the conclusion of implication~\eqref{eq:player2}, and thus the implication as well;
		\item the formula $\mood \implies (\play\land\enjoy)$ is an upper bound for $\calt_p$, being its necessary condition, what  is evident from the transitivity of implication applied to~\eqref{eq:player1} and~\eqref{eq:player2}. \done
	\end{itemize}
\end{itemize}
\end{example}

\subsection{Tightest Abstractions}

A particularly interesting class of \abstr{}s, from the point of view of maintaining the greatest inferential strength, is the class of tightest ones. The following definition formally introduces the tightest abstraction as an \abstr whose components are the tightest bounds -- that is, constituting the tightest approximations.

\begin{definition}[Tightest abstraction]\label{def:abstr-tightest}
By a \emph{tightest abstraction from a source theory $\calt_S$ with respect to a bridging theory $\calt_B$ over a vocabulary of the abstract language $\calv_A$} we mean the tightest approximation of $\calt_S$  under theory $\calt_B$ over vocabulary $\calv_A$.\done

\end{definition}

Though the lower bound of the tightest \abstr is referred to as the weakest theory, it is important to note that it is, in fact, the strongest in terms of inferential capabilities when sufficient conditions are concerned. Crucially, it encapsulates all other sufficient conditions over a given sub-language.

When $\calt_S$, $\calt_B$ and $\calv_A$ are provided as the input, the tightest \abstr{s} are uniquely determined, as stated in the following lemma.

\begin{lemma}\label{lemma:tightest}
Let $\calt_S$ be a source theory, $\calt_B$ be a bridging theory and $\calv_A$ be the vocabulary of the abstract language. Then 
$\cala\defeq\Tuple{\wsc{\calt_S}{\calt_B}{\calv_A}, \snc{\calt_S}{\calt_B}{\calv_A}}$ 
is the  tightest abstraction from the source theory $\calt_S$ with respect to $\calt_B$ on  $\calv_A$.
\end{lemma}

\proof
By Proposition~\ref{prop:abstr-approx-th}, $\cala$ is an approximate theory under $\calt_B$. By  Lemma~\ref{lemma:sncwsctapprox}, we have that \Tuple{\wsc{\calt_S}{\calt_B}{\calv_A}, \snc{\calt_S}{\calt_B}{\calv_A}} is the tightest approximation of  $\calt_S$ under $\calt_B$ over vocabulary $\calv_A$. By Definition~\ref{def:abstr-tightest}, $\cala$ is therefore the tightest abstraction from the source theory $\calt_S$ with respect to $\calt_B$ on  $\calv_A$.
\done

\begin{example}[Example~\ref{ex:game-mood} continued]\rm
The tightest abstraction from theory given by the conjunction of~\eqref{eq:player1} and~\eqref{eq:player2}, 
under bridging theory \eqref{eq:playerbridge}, when the propositional variables $g_1$ and $g_2$ are abstracted away, is given by:
\begin{itemize}
	\item the lower bound: $\lneg \mood\land (\lneg \game\lor (\play\land \enjoy))$;
	\item the upper bound: $(\mood\implies(\play\land \enjoy))\land (\mood\implies \game)$.
\end{itemize}
To see this, it suffices to use Lemma~\ref{lemma:tightest}, Lemma~\ref{lemma:wsc-snc-so}, Equations~\eqref{eq:exists} and~\eqref{eq:forall}, along with some propositional tautologies that allow for simple, equivalence-pre\-serv\-ing transformations of formulas.\done
\end{example}

Using Lemmas~\ref{lemma:tightest}, \ref{lemma:wsc-snc-so} and definitions of second-order quantifiers~\eqref{eq:exists} and~\eqref{eq:forall}, we have the following proposition.

\begin{proposition}\label{prop:tight-expr-proplog}
For any source theory $\calt_S$, bridging theory $\calt_B$ and abstract vocabulary $\calv_A$, the bounds of the tightest \abstr are expressible in classical propositional logic. \done 
\end{proposition}

\subsection{Exact Abstractions}

Another important type of \abstr{}s is exact abstractions, where lower and upper bounds are equivalent, as defined below.

\begin{definition}[Exact abstraction]\label{def:exact}
We say that an \abstr $\Tuple{\calt^l,\calt^u}$ with respect to the bridging theory $\calt_B$  is \emph{exact}, when $\calt_B\models \calt^l\equiv \calt^u$.\done
\end{definition}

We have the following  obvious proposition.

\begin{proposition}\label{prop:exact-tightest}
For arbitrary source theory $\calt_S$, bridging theory $\calt_B$ and abstract vocabulary $\calv_A$, the exact abstraction from $\calt_S$ with respect to $\calt_B$ on $\calv_A$ is the tightest abstraction from $\calt_S$ with respect to $\calt_B$ on $\calv_A$.\done 
\end{proposition}

The following example illustrates the concept of exact abstraction.

\begin{example}\label{ex:engine}\rm
Consider a simple source theory stating when an engine of a hybrid vehicle can be started, where \bc stands for ``battery charged'', \ef -- for ``engine fueled'', and \ecs -- for ``engine can start'':
\begin{equation}\label{eq:batteryfuel}
	(\bc\lor \ef)\implies \ecs.
\end{equation}
Assume one wants to abstract from detailed reasons \bc, \ef and use ``sufficient power'', \spp, instead. The bridging theory can be:
\begin{equation}\label{eq:batteryfuel-bridge}
	\spp\equiv(\bc\lor \ef).
\end{equation}
According to Proposition~\ref{prop:exact-tightest}, any exact abstraction is the tightest one, so using Lemma~\ref{lemma:tightest}, we examine the tightest \abstr from the source theory~\eqref{eq:batteryfuel} with respect to bridging theory~\eqref{eq:batteryfuel-bridge} on abstract vocabulary $\calv=\{\spp,\ecs\}$. In this case the tightest \abstr is 
$\Tuple{\wsc{\eqref{eq:batteryfuel}}{\eqref{eq:batteryfuel-bridge}}{\calv},\snc{\eqref{eq:batteryfuel}}{\eqref{eq:batteryfuel-bridge}}{\calv}}$ so, by Lemma~\ref{lemma:wsc-snc-so},
\[\Tuple{\forall \bc\forall\ef\big(\eqref{eq:batteryfuel-bridge}\implies\eqref{eq:batteryfuel}\big),\;
	\exists \bc\exists\ef\big(\eqref{eq:batteryfuel-bridge}\land\eqref{eq:batteryfuel}\big)}.\]
The lower bound is then
$\forall \bc\forall\ef\big((\spp\equiv(\bc\lor \ef))\implies ((\bc\lor \ef)\implies\ecs)\big)$,
and the upper bound is 
$\exists \bc\exists\ef\big((\spp\equiv(\bc\lor \ef))\land ((\bc\lor \ef)\implies\ecs)\big)$.
It can be easily verified that, after applying Equations~\eqref{eq:exists} and~\eqref{eq:forall}, together with some propositional tautologies to simplify the formulas, the considered tightest \abstr is
\Tuple{\spp \implies \ecs, \spp \implies \ecs}, and it is exact.\done 
\end{example} 

To highlight an idea which will be used in Theorem~\ref{thm:exact}, consider again the source theory specified by~\eqref{eq:player1} and~\eqref{eq:player2}, however with the bridging theory expressed by equivalence rather than implication:
\begin{equation}\label{eq:playerbridgeexact}
\game\equiv(g_1\lor g_2).
\end{equation}
Using calculations analogous to the previous ones, one can verify that both bounds of the tightest \abstr with respect to bridging theory~\eqref{eq:playerbridgeexact} now become equivalent to:
\[\big(\mood\implies\game\big)\land\big(\game\implies(\play\land\enjoy)\big),\] 
so the obtained abstraction is exact.

Here, and in Example~\ref{ex:engine}, the bridging theory consists of an equivalence that defines a~new concept, where the definiens is an expression used in the source theory in the same form. These examples will serve as a pattern for a generalization used in the theorem below, which provides a sufficient condition for an \abstr to be exact. 

By \emph{definitions of new concepts $c_1,\ldots,c_k$ ($k\geq 1$) over a vocabulary \calv} we mean the conjunction of equivalences:
\begin{equation}\label{eq:definitions}
(c_1\equiv A_1)\land\ldots\land  (c_k\equiv A_k),
\end{equation}
where  $\calv$ is the set of all propositional variables occurring in $A_1, \ldots,A_k$,  $c_1,\ldots,c_k\not\in\calv$ are new propositional variables such that for $1\leq i\not=j\leq k, c_i\not=c_j$. 

Let \calv be a vocabulary such that $\calv\cap\{c_1,\ldots,c_k\}=\emptyset$, and $A$ be a formula over \calv. Further, let $\mathbb{C}=\{D_i\mid i=1,\ldots,m\}$ ($m\geq 1$) be a set of subformulas of $A$ that contain only variables from $\calv'$, and such that each $D_i$ is equivalent to at least one of $A_1,\ldots,A_k$. We say that $\mathbb{C}$  \emph{covers $\calv'$ in $A$ with respect to  definitions~\eqref{eq:definitions}} if for every propositional variable $p\in\calv'$, every occurrence of $p$ in $A$ is contained in  $D_i$ for at least one $i\in\{1,\ldots,m\}$.
In such a case we also say that $\mathbb{C}$  \emph{is a cover of $\calv'$ in $A$ with respect to  definitions~\eqref{eq:definitions}}. 
The cover $\mathbb{C}$ is \emph{proper} if the subformulas in $\mathbb{C}$ are \emph{non-overlapping}, i.e., they are strictly disjoint subformulas of $A$.

\begin{example}\rm Consider $\mathbb{C}_1$ and $\mathbb{C}_2$ consisting of single formulas:
\begin{itemize}
	\item $\mathbb{C}_1\!\defeq\!\{\bc\, \lor\, \ef\}$ covers the vocabulary $\{\bc,\ef\}$ in the formula~\eqref{eq:batteryfuel} with respect to definition~\eqref{eq:batteryfuel-bridge};
	\item $\mathbb{C}_2\!\defeq\!\{g_1 \lor g_2\}$ covers the vocabulary $\{g_1,g_2\}$ in the conjunction~$\eqref{eq:player1} \land\eqref{eq:player2}$ with respect to definition~\eqref{eq:playerbridgeexact}.
\end{itemize}
Since $\mathbb{C}_1$ and $\mathbb{C}_2$ are singletons, they are obviously proper covers.
\done
\end{example}

The following theorem provides a sufficient condition for \abstr{}s to be exact.

\begin{theorem}\label{thm:exact}
Let $\calt_S$ be a source theory over vocabulary $\calv_S$ and let $\calv_A$  be an abstract vocabulary such that $\{c_1,\ldots,c_k\}\subseteq \calv_A$. Let further $\calt_B$ be a bridging theory of the form~\eqref{eq:definitions}, consisting of definitions of $c_1,\ldots,c_k$   over vocabulary $\calv_S$ such that there is a proper cover $\mathbb{C}$ of $\calv_S\setminus\calv_A$ in $\calt_S$ with respect to definitions $\calt_B$. Then the tightest \abstr from $\calt_S$ with respect to $\calt_B$ over $\calv_A$ is exact.
\end{theorem}

\proof
According to Proposition~\ref{prop:exact-tightest}, any exact abstraction is also the tightest one. By Lemma~\ref{lemma:tightest}, the tightest abstraction is
$\Tuple{\wsc{\calt_S}{\calt_B}{\calv_A}, \snc{\calt_S}{\calt_B}{\calv_A}}$. 

Let $\calv\defeq\calv_S\setminus\calv_A$ consist of all propositional variables to be abstracted away.  By Lemma~\ref{lemma:wsc-snc-so}, the tightest \abstr is $\Tuple{\forall  \calv\big(\calt_B\implies \calt_S\big), \exists  \calv\big(\calt_B\land  \calt_S\big)}$. By Definition~\ref{def:exact}, we then have to show that: 
\begin{equation}\label{eq:thmexact1}
\calt_B\models\forall  \calv\big(\calt_B\implies \calt_S\big)\equiv \exists  \calv\big(\calt_B\land  \calt_S\big).
\end{equation}
Let $\calt_S'$ denote a theory obtained  by substituting each $D_i\in\mathbb{C}$ in $\calt_S$ by a $c_j$\break  ($j\in\{1,\ldots,k\}$) such that $D_i$ is equivalent to $A_j$. It can be done consistently, since $\mathbb{C}$ is a proper cover of $\calv$ in $T_S$ with respect to~\eqref{eq:definitions}. Now, proving~\eqref{eq:thmexact1} is equivalent to showing that:
\begin{equation}\label{eq:thmexact2}
\calt_B\models\forall  \calv\big(\calt_B\implies \calt_S'\big)\equiv \exists  \calv\big(\calt_B\land  \calt_S'\big).
\end{equation}
Notice that $\calt_S'$ does not contain variables from \calv, so~\eqref{eq:thmexact2} can be equivalently rewritten as:
\begin{equation}\label{eq:thmexact3}
\calt_B\models\big(\exists  \calv(\calt_B)\implies \calt_S'\big)\equiv \big(\exists  \calv(\calt_B)\land  \calt_S'\big).
\end{equation}
Since $\calt_B\models  \exists  \calv(\calt_B)$,~\eqref{eq:thmexact3} reduces to $\calt_B\models \calt_S'\equiv  \calt_S'$, 
being trivially true.
\done

\mbox{}

That is, to apply Theorem~\ref{thm:exact}, we need to identify patterns that encompass all occurrences of the propositional variables abstracted away. Thus, the theorem is also related to abstraction patterns commonly employed in everyday reasoning.

\subsection{Layered Abstractions}

Layered abstraction is a well-known concept that allows one to simplify complex systems by organizing them into hierarchical layers. Each layer is constructed on the basis of lower layers, abstracting away selected details. In this section we show how it is embedded in the bridge-and-bound framework.

Up to now, we have considered abstraction from a given source theory. However, the result of abstraction is not a single theory, but rather a pair of theories representing approximation bounds. Consequently, to construct layered abstractions, one has to consider approximate theories as source systems. That is, the source is now an \abstr $\tuple{\calt^l,\calt^u}$ and the aim is to abstract further, as formalized in the following definition.\footnote{Since any theory $\calt$ is considered as a pair of bounds $\tuple{\calt,\calt}$, abstraction in the sense of Definition~\ref{def:a-abstr} is a special case of the layered approach.}

\begin{definition}[Layered abstraction]\label{def:abstr-layered}
By a \emph{layered abstraction}  \emph{from the source \abstr $\Tuple{\calt_S^l,\calt_S^t}$ with respect to the bridging theory $\calt_B$ on vocabulary $\calv_A$} we mean any \abstr $\tuple{\calt^l, \calt^u}$, where $\calt^l$ and $\calt^u$, respectively called its \emph{lower} and \emph{upper bound}, are theories over $\calv_A$ such that:
\begin{itemize}
	\item $\calt^l$ preserves the entailment by sufficient conditions over $\calv_A$ under theory $\calt_B$; that is, for any formula $C$ on the vocabulary $\calv_A$,
	$\calt_B\models C\implies \calt^l$ implies $\calt_B\models C\implies \calt_S^l$;
	\item $\calt^u$ preserves the entailment of necessary conditions over $\calv_A$ under theory $\calt_B$; that is, for any formula $D$ on the vocabulary $\calv_A$, $\calt_B\models \calt^u\implies D$ implies $\calt_B\models \calt_S^u\implies D$.
	\done
\end{itemize}
\end{definition}

As in Definition~\ref{def:abstr-tightest}, one can define the tightest layered abstraction as follows.

\begin{definition}[Tightest layered abstraction]\label{def:abstr-tightest-layered}
By a \emph{tightest layered abstraction from a~source \abstr $\Tuple{\calt_S^l,\calt_S^u}$ with respect to bridging theory $\calt_B$ over a vocabulary of the abstract language $\calv_A$}, we mean the \abstr $\tuple{\calt^l,\calt^u}$, for which the following conditions hold:
\begin{itemize}
	\item the \emph{lower bound}, $\calt^l$, is the weakest (with respect to $\implies$) theory over vocabulary $\calv_A$ that preserves the entailment by sufficient conditions on $\calv_A$ under $\calt_B$;
	that is, for any formula $C$ on the vocabulary $\calv_A$, $\calt_B\models C\implies \calt^l_S$  iff $\calt_B\models C\implies \calt^l;$
	\item the \emph{upper bound}, $\calt^u$, is the strongest (with respect to $\implies$) theory over vocabulary $\calv_A$ that preserves the entailment of necessary conditions on $\calv_A$ under $\calt_B$; that is,  for any formula $D$ on the vocabulary $\calv_A$, $\calt_B\models  \calt^u_S \implies D$  iff $\calt_B\models \calt^u\implies D.$ \done 
\end{itemize}
\end{definition}

By analogy to Lemma~\ref{lemma:tightest}, we have the following lemma.

\begin{lemma}\label{lemma:layeterdwscsnc}
Let $\cals=\Tuple{\calt_S^l,\calt_S^u}$ be a source \abstr, $\calt_B$ be a bridging theory and $\calv_A$ be the vocabulary of the abstract language. Then: 
\[\cala\defeq\Tuple{\wsc{\calt_S^l}{\calt_B}{\calv_A}, \snc{\calt_S^u}{\calt_B}{\calv_A}}\] 
is the tightest \abstr from \cals with respect to $\calt_B$ on  $\calv_A$.
\done
\end{lemma}

We also have the following theorem, which demonstrates that the tightest layered abstractions are compositional.

\begin{theorem}\label{thm:layered} 
Let $\calv_S, \calv_A', \calv_A''$ be vocabularies such that $\calv_S\cap\calv_A''=\emptyset$.\footnote{The assumption that $\calv_S\cap\calv_A''=\emptyset$ reflects a natural methodological principle that symbols rejected at lower layers as being too detailed are not reused at more abstract layers.} Let further \break 
$\cals=\Tuple{\calt_S^l,\calt_S^u}$ be a source \abstr over vocabulary $\calv_S$, $\calt_B'$ be a bridging theory over $\calv_S\cup\calv_A'$, $\calt_B''$ be a bridging theory over $\calv_A'\cup\calv_A''$, and:
\begin{itemize}
	\item $\cala'\defeq\Tuple{{\calt_S^l}',{\calt_S^u}'}\defeq\Tuple{\wsc{\calt_S^l}{\calt_B'}{\calv_A'}, \snc{\calt_S^u}{\calt_B'}{\calv_A'}}$ be the tightest \abstr from \cals with respect to $\calt_B'$ on $\calv_A'$;
	\item $\cala''
	\defeq\Tuple{\wsc{{\calt_S^l}'}{\calt_B''}{\calv_A''}, \snc{{\calt_S^u}'}{\calt_B''}{\calv_A''}}$ be the tightest \abstr from $\cala'$ with respect to $\calt_B''$ on $\calv_A''$.
\end{itemize}  
Then $\cala''$ is the tightest \abstr from \cals with respect to $\calt_B'\land\calt_B''$ on $\calv_A''$, i.e.,
\begin{equation}\label{eq:compositional}
	\cala''=\Tuple{\wsc{{\calt_S^l}}{\calt_B'\land\calt_B''}{\calv_A''}, \snc{{\calt_S^u}}{\calt_B'\land\calt_B''}{\calv_A''}}.
\end{equation}
\end{theorem}

\proof Notice that the formulas appearing in~\eqref{eq:compositional} are built over vocabulary\break $\calv_S\cup\calv_A'\cup\calv_A''$. By assumption,  $\calv_S\cap\calv_A''=\emptyset$, so:
\begin{equation}\label{eq:setminus}
(\calv_S\cup\calv_A'\cup\calv_A'')\setminus \calv_A''= (\calv_S\setminus \calv_A')\cup(\calv_A'\setminus \calv_A'').
\end{equation}

\begin{itemize}
\item Let us first prove that $\wsc{{\calt_S^l}}{\calt_B'\land\calt_B''}{\calv_A''}\equiv\wsc{{\calt_S^l}'}{\calt_B''}{\calv_A''}$. 

By Lemma~\ref{lemma:wsc-snc-so}, $\wsc{{\calt_S^l}}{\calt_B'\land\calt_B''}{\calv_A''}  \equiv  
\forall ((\calv_S\cup\calv_A'\cup\calv_A'')\setminus \calv_A'')\big((\calt_B'\land\calt_B'')\implies \calt_S^l\big)$. By~\eqref{eq:setminus}, this formula is equivalent to: 
\[
\begin{array}{l}
	\forall ((\calv_S\setminus \calv_A')\cup(\calv_A'\setminus \calv_A''))\big((\calt_B'\land\calt_B'')\implies \calt_S^l\big)\equiv\\
	\qquad \forall (\calv_S\setminus \calv_A')\,\forall(\calv_A'\setminus \calv_A'')\big((\calt_B'\land\calt_B'')\implies \calt_S^l\big)\equiv\\
	\qquad \forall (\calv_S\setminus\calv_A')\,\forall(\calv_A'\setminus\calv_A''))\big(\calt_B''\implies (\calt_B'\implies\calt_S^l)\big).
\end{array}	
\]

Of course, $\calt_B''$ contains no symbols from $\calv_S \setminus \calv_A'$, so the last formula is equivalent to:
\[
\begin{array}{l}
	\forall(\calv_A'\setminus\calv_A'')\big(\calt_B''\implies \underbrace{\forall(\calv_S\setminus\calv_A')(\calt_B'\implies \calt_S^l)}_{\wsc{\calt_S^l}{\calt_B'}{\calv_A'}\bydefequiv {\calt_S^l}'}\big)\equiv \\
	\qquad \forall(\calv_A'\setminus\calv_A'')\big(\calt_B''\implies  {\calt_S^l}'\big)\equiv  \wsc{{\calt_S^l}'}{\calt_B''}{\calv_A''}.
\end{array} 
\]

Thus, indeed, $\wsc{{\calt_S^l}}{\calt_B'\land\calt_B''}{\calv_A''}\equiv \wsc{{\calt_S^l}'}{\calt_B''}{\calv_A''}$.

\item It remains to prove that $\snc{{\calt_S^u}}{\calt_B'\land\calt_B''}{\calv_A''}\equiv\snc{{\calt_S^u}'}{\calt_B''}{\calv_A''}$. 

By Lemma~\ref{lemma:wsc-snc-so}, $\snc{{\calt_S^u}}{\calt_B'\land\calt_B''}{\calv_A''}  \equiv  
\exists ((\calv_S\!\cup\!\calv_A'\!\cup\!\calv_A'')\setminus \calv_A'')\big((\calt_B'\!\land\!\calt_B'')\!\land\! \calt_S^u\big)$. \break By~\eqref{eq:setminus}, this formula is equivalent to: 
\[
\!\!\!\!\!\!\!\!\!\!\!\!\begin{array}{l}
	\exists ((\calv_S\!\setminus\! \calv_A')\cup(\calv_A'\!\setminus\! \calv_A''))\big(\calt_B'\!\land\!\calt_B''\!\land\! \calt_S^u\big)\equiv
	\exists (\calv_S\!\setminus\! \calv_A')\,\exists(\calv_A'\!\setminus\! \calv_A'')\big(\calt_B''\!\land\!\calt_B'\!\land\! \calt_S^u\big).
\end{array}	
\]
Since $\calt_B''$ contains no symbols from $\calv_S \setminus \calv_A'$,
\[
\!\!\!\!\!\!\!\!\!\!\!\!\begin{array}{l}
	\exists (\calv_S\!\setminus\! \calv_A')\,\exists(\calv_A'\!\setminus\! \calv_A'')\big(\calt_B''\!\land\!\calt_B'\!\land\! \calt_S^u\big)\equiv  \exists(\calv_A'\!\setminus\!\calv_A'')\big(\calt_B''\land \underbrace{\exists(\calv_S\!\setminus\!\calv_A')(\calt_B'\!\land\! \calt_S^u)}_{\snc{\calt_S^u}{\calt_B'}{\calv_A'}\bydefequiv {\calt_S^u}'}\big)\equiv\\
	\qquad \exists(\calv_A'\!\setminus\!\calv_A'')\big(\calt_B''\!\land\! {\calt_S^u}'\big)\equiv \snc{{\calt_S^u}'}{\calt_B''}{\calv_A''}.
\end{array} 
\]
Thus, indeed, $\snc{{\calt_S^u}}{\calt_B'\land\calt_B''}{\calv_A''}\equiv \snc{{\calt_S^u}'}{\calt_B''}{\calv_A''}$.
\done
\end{itemize}

\noindent Using Theorem~\ref{thm:layered}, one can easily see that:
\begin{itemize}
\item tightest layered abstractions can actually be seen as a single abstraction combining all layers;
\item given that the original bridging theory $\calt_B$ is a conjunction of sub-theories, the complex single tightest abstraction with respect to $\calt_B$ can serve as a hint for how the original abstraction might be structured into smaller layers, potentially easier to compute and explain, using the sub-theories as bridging theories for the sub-layers.
\end{itemize}

\subsection{Remarks}

We have started defining our bridge-and-bound framework from the loosest concept of \abstr, assuming only that sufficient and necessary conditions are preserved. Although this permits even trivial abstractions, such as $\tuple{\false,\true}$, the rationale is that the tightest or exact abstractions are not always desirable, particularly when reasoning complexity is a concern and one may wish to approximate theories with simpler, preferably tractable ones~\cite{CadoliD97,DarwicheM24,kautzselmanK96}, rather than necessarily fighting for the tightest. 

While sufficient conditions are not considered as frequently as necessary ones, they play a~crucial role in several important contexts, including abduction, explanation, and action theories, among many others. Their use is essential for reasoning about what guarantees the occurrence of certain outcomes, for constructing explanatory frameworks, and for understanding the effects of actions in dynamic systems. For detailed discussions and applications, see, e.g.,~\cite{dlss2006a,dlssnc,dsz-forgetting-aij,FengCTL,FengMU,KoopmannDTS20,Lin,Lin01,liu-lakemayer}.

Tightest \abstr are important because they provide the strongest inferential bounds that, given the abstract vocabulary and bridging theory, preserve both sufficient and necessary conditions. Exact \abstr{s} are \emph{ideal} in the sense that these bounds are identical modulo equivalence. Although this requirement may seem very strong, it can be achieved in many practical cases. For example, in Theorem~\ref{thm:exact} we provide a sufficient condition for exactness when definitional bridging theories are used. The underlying idea of this theorem can also be applied to automate the process of finding \abstr by looking at repetitions of subformulas in the source theories. The idea is well grounded in the intuition that humans identify abstractions by discovering patterns and encapsulating them as new concepts. These patterns are formally captured as definientia in definitions~\eqref{eq:definitions} (and in the first-order case considered in the next section, in~\eqref{eq:definitions-fo}). To automate the process, one searches for recurring patterns (modulo equivalence) and encapsulate them as new concepts. This form of abstraction is common in everyday life, where one observes patterns, e.g., ``barks, has tail, and four legs'', and uses a general concept, such as ``dog'', rather than enumerating all the detailed characteristics. The same idea is omnipresent in science and engineering. For example, in programming, whenever a piece of code is repeated, it is typically encapsulated in a function (method or procedure) and used as an abstraction that is easier to understand and maintain. It would be interesting to identify other sufficient conditions for exactness, and possibly necessary ones, that are more abstract than those directly given in~Definition~\ref{def:exact}.

Layered abstractions are very interesting from methodological and pragmatic viewpoints, because they allow us to organize complexity into manageable pieces. Methodologically, layering helps us separate concerns: each layer can focus on a~specific level of representation or function, which makes reasoning, verification, and understanding much easier. Pragmatically, this approach facilitates reuse, modularity, and adaptability. When we change a~layer, we can often keep the other layers intact, reducing the cost of maintenance and evolution. Layered abstractions also enable incremental understanding: one can analyze the system or model layer by layer instead of being overwhelmed by its full complexity at once. This combination of conceptual clarity and practical flexibility is one of the main factors that make layered designs so widely used across fields such as software engineering or scientific modeling. Theorem~\ref{thm:layered} provides a~pragmatic tool for identifying layers. It can be implemented to assist theorists in discovering the most intuitive and useful layers.

\section{\!\!\!Extending\,the\,Framework\,to\,the\,First-Order~Case}\label{sec:fo}

\subsection{First- and Second-Order Logic}

Let us now focus on the case when source and bridging theories are expressed in first-order logic.
To define the first-order language, we assume finite sets:
\begin{itemize}
\item \calc: consisting of \emph{constants} representing domain elements;
\item $\calv_I$: consisting of  \emph{(first-order) variables} ranging over domain elements;
\item $\calr$: consisting of \emph{relation symbols}.
\end{itemize}
In addition to propositional connectives, \emph{universal} and \emph{existential quantifiers} $\forall, \exists$ ranging over the domain are allowed.\footnote{Function symbols are avoided, as is standard in rule-based query languages and underlying belief bases~\cite{AHV}, what simplifies the presentation of complexity results, discussed in Section~\ref{sec:compaspects}. Of course, functions can be defined by relations in a standard manner.}  A variable occurrence $x$ is \emph{bound} in a formula $A$ when it occurs in the scope of a~quantifier $\forall x$ or $\exists x$. A variable occurrence is \emph{free} when the occurrence is not bound. In the rest of the paper, $A(\bar{x})$ is used to indicate that $\bar{x}$ are all variables that occur free in $A$. Formula $A$ is \emph{closed} when it contains no free variables.

A \emph{first-order theory (belief base)} is a finite set of first-order formulas,  understood as a~single formula formed, like in the propositional case, by the conjunction of the formulas it contains. A~first-order theory is \emph{closed} if it consists of closed formulas only. 

The semantics for first-order logic is standard, assuming:
\begin{itemize}
\item a \emph{domain}: a non-empty set whose elements are values assigned to constants and variables;
\item \emph{relations}: (possibly empty) sets consisting of tuples of domain elements, interpreting relation symbols.
\end{itemize} 
Quantifiers are given their traditional interpretation: 
\begin{itemize}
\item $\forall x A(x)$: every element $d$ in the domain satisfies $A(x)$ when $x=d$;
\item $\exists x A(x)$: there exists an element $d$ in the domain such that $A(x)$ is satisfied when $x=d$.
\end{itemize}

Second-order logic is needed for characterizing weakest sufficient and strongest necessary conditions of first-order formulas. For second-order logic we assume the language of first-order logic extended with \emph{second-order variables}, $\calv_{\mathit{II}}$, ranging over relations, and \emph{second-order quantifiers} $\forall X, \exists X$ quantifying over relations, with the standard meaning:
\begin{itemize}
\item $\forall X A(X)$: every relation $R$ consisting of tuples of domain elements satisfies $A(X)$ when $X=R$;
\item $\exists X A(X)$: there exists a relation $R$ consisting of tuples of domain elements such that  $A(X)$ is satisfied when $X=R$.
\end{itemize}

There is a vast body of literature on first- and second-order logic. For a detailed review, see, e.g.,~\cite{gss,mendelson,sep-logic-higher-order}.

\subsection{First-Order \Abstr{}s}

Having defined the language of first-order logic, lifting \abstr to the first-order counterparts is immediate. That is, in Definitions~\ref{def:a-abstr}, \ref{def:abstr-tightest}, \ref{def:exact}, \ref{def:abstr-layered} and~\ref{def:abstr-tightest-layered} one simply has to assume that:
\begin{itemize}
\item the underlying theories are first-order; however, we always assume the bridging theories to be closed;\footnote{Only bridging theories need to be closed, as only they are used in applying the deduction theorem for first-order logic, which is needed to prove the first-order version of Lemma~\ref{lemma:wsc-snc-so}, as given in~\cite{dlssnc}.}
\item the vocabularies consist of relation symbols rather than propositional variables.
\end{itemize} 
Similarly, the results expressed as Lemma~ \ref{lemma:sncwsctapprox}, Propositions~\ref{prop:abstr-approx-th}, \ref{prop:exact-tightest}, Lemmas~\ref{lemma:tightest}, \ref{lemma:layeterdwscsnc} and Theorems~\ref{thm:exact},~\ref{thm:layered} can easily be lifted to the first-order case.\footnote{Note also that the first-order version of Lemma~\ref{lemma:wsc-snc-so}, which is needed in the first-order version of Lemma~\ref{lemma:tightest}, is formulated in~\cite[Lemma 4.1]{dlssnc}.} In the case of Theorem~\ref{thm:exact}, the bridging theory consists of definitions of relations, involving first-order variables, that is, rather than~\eqref{eq:definitions}, one considers:
\begin{equation}\label{eq:definitions-fo}
\big(c_1(\bar{x}_1)\equiv A_1(\bar{x}_1)\big)\land\ldots\land  \big(c_k(\bar{x}_k)\equiv A_k(\bar{x}_k)\big),
\end{equation}
where $\bar{x}_1, \ldots,\bar{x}_k$ are all free variables occurring respectively in $A_1(\bar{x}_1),\ldots, A_k(\bar{x}_k)$.

On the other hand, although the tightest propositional  \abstr is expressible in classical propositional logic (see Proposition~\ref{prop:tight-expr-proplog}), this property does not carry over to the first-order case, as shown below.

\begin{proposition}\label{prop:fo-tight}
There are source and bridging theories $\calt_S$, $\calt_A$ and an abstract vocabulary $\calv_A$ such that the bounds of the tightest first-order \abstr from $\calt_S$ with respect to $\calt_B$ over $\calv_A$ are not expressible as classical first-order theories.
\end{proposition}

\proof
Consider $\calt_S$ consisting of first-order induction axiom over the standard language of arithmetics together with a relation symbol $p$:\footnote{Note that $s(x,y)$ stands for ``$y$ is a successor of $x$'', and~\eqref{eq:fo-ind} is a formula involving the relation symbol $p$ ranging over \emph{all} relations, rather than the Peano induction schema in which $p$ would range over arbitrary first-order formulas, not representing \emph{all} relations.}
\begin{equation}\label{eq:fo-ind}
\calt_S\defequiv p(0)\land \forall x\forall y\big((p(x)\land s(x,y))\implies p(y)\big).
\end{equation}
Let $T_B\defequiv\true$, and the abstract vocabulary $\calv_A$ consist of all relation symbols excluding $p$. Then, the lower bound of the tightest \abstr from $\calt_S$ with respect to $\calt_B$ over $\calv_A$ is:
\begin{equation}\label{eq:so-ind}
\wsc{\calt_S}{\calt_B}{\calv_A}\equiv \forall p\big(p(0)\land \forall x\forall y\big((p(x)\land s(x,y))\implies p(y)\big)\big),
\end{equation}
being the second-order induction axiom known  to be inexpressible in classical first-order logic (see, e.g.~\cite{sep-logic-higher-order}).

A similar argument applies to show that the upper bound may not be expressible in classical first-order logic, however, this time with $\calt_S$ defined as the negation of~\eqref{eq:fo-ind}.

\done

The following example illustrates the first-order bridge-and-bound framework.

\begin{example}\label{ex:fo}\rm
Consider a scenario in which the average speed of vehicles is measured over a road segment. When a car enters the monitored segment, a timestamp is recorded. Upon leaving, the travel time is calculated and the system reports speed limit violations. We will use the constants $\lgh, \slm$ and $\mnt$, where $\lgh$ is the length of the road segment, $\slm$ is the speed limit over the road segment,  and $\mnt$ is the minimum travel time over the segment to comply with the speed limit. Of course, $\slm=\lgh/\mnt$, which will allow us to make suitable substitutions for better clarity of the resulting formulas.\footnote{To simplify the calculations, we will freely apply the laws of arithmetic without explicitly listing them within the considered bridging theory.}

The vocabulary of the source system includes constants $\lgh, \mnt$ and relation symbols $\tim, \rep$, where:
\begin{itemize}
	\item $\tim(c,t)$ states that car $c$ traveled the road segment in time $t$;  
	\item $\rep(c)$ states that car $c$ is reported as having traveled the road segment in a~time shorter than, say, $0.8 * \mnt$.
\end{itemize}
The source theory $\calt_S$ in this case is:\footnote{We assume that  standard arithmetical operations over rationals are within the source and abstract vocabulary.}
\begin{equation}\label{eq:road-s}
	\forall c\forall t\Big(\big(\tim(c,t)\land\, t< 0.8*\mnt\big) \implies \rep(c)\Big).
\end{equation}
At a more abstract level, one is interested in average speed rather than travel time, as a more common and informative measure. The abstract vocabulary $\calv_A$ excludes $\tim$ while including  new relation symbol $\vel(c,v)$ indicating that the average speed of car $c$ on the road segment is $v$. The bridging theory $\calt_B$ may then connect time and speed, e.g, by:
\begin{equation}\label{eq:road-b}
	\forall c\forall t\big(\tim(c,t) \implies \vel(c,\lgh/t)\big)\land
	\forall c\forall v	\big(\vel(c,v)\implies \tim(c,\lgh/v)\big).
\end{equation}
By the first-order version of Lemma~\ref{lemma:tightest}, the tightest \abstr from $\calt_S$ with respect to $\calt_B$ over $\calv_A$ is   $\Tuple{\wsc{\eqref{eq:road-s}}{\eqref{eq:road-b}}{\calv_A}, \snc{\eqref{eq:road-s}}{\eqref{eq:road-b}}{\calv_A}},
$
i.e.,\footnote{For simplicity, slightly abusing notation we use the same notation for relation symbols and corresponding second-order variables.}
\begin{equation}\label{eq:road-tightest-so}
	\Tuple{\forall\tim\big(\eqref{eq:road-b}\implies \eqref{eq:road-s}\big),\; 
		\exists\tim\big(\eqref{eq:road-b}\land \eqref{eq:road-s}\big)
	}.
\end{equation}
\done
\end{example}

The pair~\eqref{eq:road-tightest-so} contains second-order quantifiers $\forall\tim,\exists\tim$ that bind second-order variables, so the definitions of propositional second-order quantifiers given in~\eqref{eq:exists} and~\eqref{eq:forall} cannot be applied. To compute first-order equivalents of~\eqref{eq:road-tightest-so}, the technique used in~\cite{dlssnc} applies Ackermann's lemma~\cite{Ack} (see also~\cite{gss} or a broad discussion in~\cite{dsz-forgetting-aij}).\footnote{In fact, a fixpoint lemma from~\cite{NS} is also used in~\cite{dlssnc}; however, the resulting formulas are expressed in fixpoint logic, and thus beyond first-order logic.} To formulate the lemma, let us remind that a formula $B$ is \emph{positive} with respect to relation symbol $r$ (second-order variable $X$)  if every occurrence of $r$ in $B$ is under an even number of negations (assuming that $C\implies D$ is understood as $\lneg C\lor D$, so implicitly introduces a negation in front of $C$). Formula $B$ is \emph{negative} with respect to relation symbol $r$ (second-order variable $X$) if every occurrence of $r$ in $B$ is under an odd number of negations.

\begin{lemma}[Ackermann~\cite{Ack}]\label{lemma:ack-fo}
Let $X$ be a second-order variable representing\break $k$-argument relations ($k\geq 0$), $A$ be a  first-order formula without occurrences of $X$, $B$ be a first-order formula perhaps involving subformulas of the form $X(e_1,\ldots,e_k)$ where $X$ is applied to expressions $e_1,\ldots,e_k$, and let $\bar{x}\defeq x_1,\ldots, x_k$ be a $k$-tuple of distinct first-order variables. Then:
\begin{itemize}
	\item  if $B$ is positive with respect to\ $X$ then:
	\begin{equation}\label{eq:ack-pos-fo}
		\exists X\big(\forall\bar{x}(X(\bar{x})\implies
		A(\bar{x}))\land B\big)\ \equiv\ B(X(\bar{x})=A(\bar{x}));
	\end{equation} 
	\item if $B$ is negative with respect to\ $X$ then:
	\begin{equation}\label{eq:ack-neg-fo}
		\exists X\big(\forall\bar{x}(A(\bar{x})\implies X(\bar{x}))\land B\big) \ \equiv\ B(X(\bar{x})=A(\bar{x})),
	\end{equation}  
\end{itemize}
where $X(\bar{x})=A(\bar{x})$ denotes the substitution of each occurrence of $X(e_1,\ldots,e_k)$ by the formula $A$ in which every occurrence of variable $x_i$ ($i=1,\ldots,k$) is replaced by $e_i$. \done
\end{lemma}
The strategy of using Lemma~\ref{lemma:ack-fo} is first to transform equivalently the input formula so that it takes the form~\eqref{eq:ack-pos-fo} or~\eqref{eq:ack-neg-fo}. If this step is successful then apply the lemma and eliminate $\exists X$. If not, the algorithm fails.\footnote{That does not mean the input formula is not equivalent to a first-order formula, only that the applied transformations were unsuccessful.} 
To deal with second-order universal quantifiers, one  applies Lemma~\ref{lemma:ack-fo} to $\exists X(\lneg A(X))$ and then negates the result, what reflects the deMorgan law $\forall X A(X)\equiv \lneg\exists X(\lneg A(X))$.

The following example illustrates the use of Lemma~\ref{lemma:ack-fo} in computing tightest \abstr.

\begin{example}[Example~\ref{ex:fo} continued]\label{ex:fo-cont}\rm
The lower bound of the considered \abstr is specified in~\eqref{eq:road-tightest-so} as $\forall\tim\big(\eqref{eq:road-b}\implies \eqref{eq:road-s}\big)$. To apply Lemma~\ref{lemma:ack-fo}, we use its De Morgan's equivalent  $\lneg\exists\tim\big(\eqref{eq:road-b}\land\lneg \eqref{eq:road-s}\big)$ and transform it to a required form:
\begin{align*}
	& \lneg\exists\tim\Big(\forall c\forall t\big(\tim(c,t) \implies \vel(c,\lgh/t)\big)\land
	\forall c\forall v	\big(\vel(c,v)\implies \tim(c,\lgh/v)\big)\land\\
	& \qquad \quad\;\; \lneg \forall c\forall t\big(\big(\tim(c,t)\land\, t< 0.8*\mnt\big) \implies \rep(c)\big)
	\Big)\equiv\\
	& \lneg\exists\tim\Big(\forall c\forall t\big(\tim(c,t) \implies \vel(c,lgh/t)\big)\land
	\forall c\forall v	\big(\vel(c,v)\implies \tim(c,lgh/v)\big)\land\\
	& \qquad \quad\;\; \exists c\exists t\big(\tim(c,t)\land\, t< 0.8*\mnt\land\lneg \rep(c)\big)
	\Big).
\end{align*}
Now the formula under the outermost negation is in the form~\eqref{eq:ack-pos-fo}. After applying Lemma~\ref{lemma:ack-fo}, we obtain the following equivalent:
\[
\begin{array}{l}
	\lneg\Big(\forall c\forall v	\big(\vel(c,v)\implies \vel(c,\lgh/(\lgh/v))\big)\land\\
	\;\;\;\;\;\exists c\exists t\big(vel(c,\lgh/t)\land\, t< 0.8*\mnt\land\lneg \rep(c)\big)
	\Big).
\end{array}
\]
Since $\lgh/(\lgh/v)=v$ the first conjunct is equivalent to \true. That is, the lower bound of the tightest \abstr, $\forall\tim\big(\eqref{eq:road-b}\implies \eqref{eq:road-s}\big)$, is equivalent to:
\begin{equation}\label{eq:roadwwsc}
	\forall c\forall t\Big(\big(\vel(c,\lgh/t)\land\, t< 0.8*\mnt\big)\implies \rep(c)\Big).
\end{equation}
One can refine the result further by introducing $v=\lgh/t$ that stands for the car's average speed. Then the formula~\eqref{eq:roadwwsc} becomes equivalent to: 
\[\forall c\forall v\Big(\big(\vel(c,v)\land\, \lgh/v< 0.8*\mnt\big)\implies \rep(c)\Big),\]
i.e., to:
\begin{equation}\label{eq:roadwwsc-prefinal}
	\forall c\forall v\Big(\big(\vel(c,v)\land\, v> 1.25*\lgh/\mnt\big)\implies \rep(c)\Big).
\end{equation}
Recall that $\slm=\lgh/\mnt$ is the maximal allowed average velocity. So, in terms of the more abstract notion of velocity, formula~\eqref{eq:roadwwsc-prefinal} states that the reported cars are those whose average velocity over the considered road segment exceeded $1.25$ times the maximal allowed average velocity, and can finally be formulated as:
\begin{equation}\label{eq:roadwwsc-final}
	\forall c\forall v\Big(\big(\vel(c,v)\land\, v> 1.25*\slm\big)\implies \rep(c)\Big).
\end{equation}

The upper bound of \abstr is specified  in~\eqref{eq:road-tightest-so} as $\exists\tim\big(\eqref{eq:road-b}\land \eqref{eq:road-s}\big)$. To apply Lemma~\ref{lemma:ack-fo}, we transform it to a required form:
\begin{align*}
	\exists\tim\Big(&\forall c\forall t\big(\tim(c,t) \implies \vel(c,\lgh/t)\big)\land
	\forall c\forall v	\big(\vel(c,v)\implies \tim(c,\lgh/v)\big)\land\\
	& \forall c\forall t\big(\big(\tim(c,t)\land\, t< 0.8*\mnt\big) \implies \rep(c)\big)
	\Big)\equiv\\
	\exists\tim\Big(&\forall c\forall t\big(\tim(c,t) \implies \vel(c,\lgh/t)\big)\land\\
	&\forall c\forall v\forall x	\big(\big(\vel(c,v)\land x=\lgh/v\big)\implies \tim(c,x)\big)\land\\
	& \forall c\forall t\big(\big(\tim(c,t)\land\, t< 0.8*\mnt\big) \implies \rep(c)\big)
	\Big).
\end{align*}
To make the calculations clear, let us exchange the order of conjunction in the  first line and move $\forall v$ inside the formula:
\begin{align*}
	\exists\tim\Big(&\forall c\forall x	\big(\exists v\big(\vel(c,v)\land x=\lgh/v\big)\implies \tim(c,x)\big)\land \\
	&\forall c\forall t\big(\tim(c,t) \implies \vel(c,\lgh/t)\big)\land\\
	& \forall c\forall t\big(\big(\tim(c,t)\land\, t< 0.8*\mnt\big) \implies \rep(c)\big)
	\Big).
\end{align*}
The formula is now in form~\eqref{eq:ack-neg-fo}. After applying Lemma~\ref{lemma:ack-fo}, we obtain its equivalent: 
\begin{align*}
	&\forall c\forall t\big(\exists v\big(\vel(c,v)\land t=\lgh/v\big) \implies \vel(c,\lgh/t)\big)\land\\
	& \forall c\forall t\big(\exists v\big(\big(\vel(c,v)\land t=\lgh/v\big)\land\, t< 0.8*\mnt\big) \implies \rep(c)\big).
\end{align*}
The first line of the above formula reduces to \true. In the second line, $t$ can be replaced by $\lgh/v$, and $\exists v$ can be moved to the prefix, becoming there $\forall v$:
\[
\forall c\forall v\big(\big(\vel(c,v)\land\, \lgh/v< 0.8*\mnt\big) \implies \rep(c)\big).
\]
One can easily observe that it ie equivalent to:
\[
\forall c\forall v\big(\big(\vel(c,v)\land\, v>1.25*\lgh/\mnt\big) \implies \rep(c)\big),
\]
i.e., to:
\begin{equation}\label{eq:roadsnc}
	\forall c\forall v\big(\big(\vel(c,v)\land\, v>1.25*\slm\big) \implies \rep(c)\big),
\end{equation}
being again related to reporting a car when its average speed exceeds a maximal allowed average speed by more over $25\%$.

Summing up, the tightest \abstr in the considered scenario is \Tuple{\eqref{eq:roadwwsc-final},\eqref{eq:roadsnc}}. Since \eqref{eq:roadwwsc-final} is the same as \eqref{eq:roadsnc}, this \abstr is exact.	\done
\end{example}

\subsection{Remarks}

The extension of the bridge-and-bound framework to first-order logic, though conceptually natural, raises important questions especially whenever second-order quantifiers are involved. It is well-known that second-order constructs radically increase the involved complexity. Therefore, whenever possible, one should struggle to eliminate second-order quantifiers. We have used Ackermann's lemma, but other techniques are also known~\cite{gss}. The problem of equivalence of second and first-order formulas is highly undecidable (even when the first-order formula is just $\true$), so none of the techniques works in general. Proposition~\ref{prop:fo-tight} formally shows that the tightest \abstr is not expressible in first-order logic. However, the techniques of eliminating second order quantifiers work in many cases known from practice of modeling~\cite{gss}. It is also interesting that second-order quantifier elimination is also known and intensively investigated in the area of forgetting~\cite{Delgrande17,dsz-forgetting-aij,Eiter-kern-isberner,Lin94forgetit}.

In Examples~\ref{ex:fo} and~\ref{ex:fo-cont}, formula~\eqref{eq:road-b} effectively defines velocity in terms of time and distance. However, this definition is somewhat implicit, relying on the standard rules for manipulating arithmetic expressions. This illustrates that an extension of Theorem~\ref{thm:exact} to the first-order setting works well when all underlying theories, such as arithmetic, are explicitly incorporated within the bridging theory. In the examples, we applied these laws as given; in practice, however, implemented techniques must be equipped to handle them formally. This highlights the importance of explicitly encoding the relevant background knowledge, so that the abstraction process can correctly account for the logical and computational properties of the underlying theories. Such an approach ensures that exact abstractions remain valid even in the presence of implicit mathematical reasoning.

We have applied Ackermann’s lemma in the first-order setting. Naturally, the lemma is also applicable to propositional theories, where it can be used to eliminate propositional variables in a systematic and sound manner. More generally, Ackermann’s lemma provides a~principled approach for reducing the complexity of a theory by replacing quantified symbols with their explicit definitions, thereby facilitating both abstraction and reasoning. Its applicability extends across different logical frameworks, making it a versatile tool in both first-order and propositional contexts.

\section{Computational Aspects}\label{sec:compaspects}

The main computational questions we address are as follows: \\
\centerline{\parbox{11.5cm}{
	\begin{itemize}
		\item[(Q1)] verifying whether a pair of theories forms an \abstr;
		\item[(Q2)] computing the tightest abstractions so that they can be expressed within the considered logic;
		\item[(Q3)] verifying whether a given \abstr is exact;
		\item[(Q4)] reasoning by entailment and by querying.
	\end{itemize}
	}}
	
	Most of the answers can be readily adapted from the literature, including the excellent and comprehensive surveys provided in books \cite{AHV,EF,goldreich,Imbook}. We will also make use of the results included in~\cite{dlssnc,gss}.
	
	\subsubsection*{\normalsize Question (Q1)} 
	
	The conditions on \abstr are formulated in Definition~\ref{def:a-abstr} in terms of universal second-order formulas. Indeed, the condition on $\calt^l$ can be expressed by: 
	\begin{equation}\label{eq:question1a}
\calt_B\models \forall C\big((C\implies \calt^l)\implies (C\implies \calt_S^l)\big),
\end{equation}
and on $\calt^u$, by:
\begin{equation}\label{eq:question1b}
\calt_B\models \forall D\big((\calt^u\implies D)\implies (\calt_S^u\implies D)\big).
\end{equation}
This shows that, in the propositional case, the problem of answering (Q1) is \conptime-complete and, in the first-order case, it is partially decidable, since verifying~\eqref{eq:question1a} and~\eqref{eq:question1b} reduces to first-order entailment, being equivalent to verifying that\break $\calt_B \models \big(C(\bar{x}) \implies \calt^l\big) \implies \big(C(\bar{x}) \implies \calt_S^l\big)$ and $\calt_B \models \big(\calt^u \implies D(\bar{y})\big) \implies \big(\calt_S^u \implies D(\bar{y})\big)$,  where $\bar{x}$ and $\bar{y}$ are tuples consisting of all free variables occurring in $\calt_S^l$ and $\calt_S^u$, respectively.

\subsubsection*{\normalsize Question (Q2)}

Notice that the tightest \abstr{s} are expressed by the weakest sufficient and strongest necessary conditions (Lemma~\ref{lemma:tightest}), which are characterized using second-order quantifiers (Lemma~\ref{lemma:wsc-snc-so}). Therefore, the question (Q2) reduces to eliminating second-order quantifiers that take the formulas outside the scope of classical propositional (first-order) logic. For a comprehensive review of second-order quantifier elimination techniques see~\cite{gss}. In~\cite{dlssnc}, Ackermann's lemma (Lemma~\ref{lemma:ack-fo} in the present paper) is used in the context of the $wsc()$ and $snc()$ operators.

In fact, when an application of Lemma~\ref{lemma:ack-fo} is possible, it may, in the worst case, increase the size of the resulting formula to $O(n^2)$, where $n$ is the size of formulas~\eqref{eq:ack-pos-fo} and~\eqref{eq:ack-neg-fo}.\footnote{Typically, the growth is linear in $n$, and in many cases the formula even becomes smaller.} The main difficulty lies in transforming the input formulas into one of the forms~\eqref{eq:ack-pos-fo} and~\eqref{eq:ack-neg-fo}. The \dls algorithm~\cite{dls}, which performs such transformations and then uses this lemma, applies to large classes of formulas. In particular, as shown in~\cite{dls}, the \dls algorithm subsumes most other known techniques for computing circumscription. Moreover, \cite{Conradie06} shows that the \dls algorithm covers all Sahlqvist formulas, an important class in modal correspondence theory. However, in the case of first-order logic, elimination of second-order quantifiers is not always possible, as shown in Proposition~\ref{prop:fo-tight}.

Second-order quantifiers can always be eliminated from propositional theories using definitions~\eqref{eq:exists} and~\eqref{eq:forall}. However, each quantifier elimination doubles the size of the formula, making the result exponential in the number of quantifiers to be eliminated. As mentioned earlier, Ackermann's lemma often performs much better, as the resulting formula may even be smaller than the input.

\subsubsection*{\normalsize Question (Q3)}

Given an \abstr $\cala = \Tuple{\calt^l, \calt^u}$ with respect to a bridging theory $\calt_B$, verifying whether $\cala$ is exact, according to Definition~\ref{def:exact}, reduces to checking whether\break $\calt_B \models \calt^l \equiv \calt^u$. In other words, one must verify that, under the assumptions of the bridging theory, the lower and upper bounds are logically equivalent. When $\cala$ is expressed in propositional logic, this verification problem is \conptime-complete, as it amounts to propositional entailment checking. In contrast, when $\cala$ is expressed in first-order logic, the problem is only partially decidable. This distinction motivates the search for restricted, tractable fragments or automated heuristics to make exactness checking feasible in practice. 

\subsubsection*{\normalsize Question (Q4)}

Given a propositional \abstr, checking whether it entails a given condition (in the case of necessary conditions) or is entailed by it (in the case of sufficient conditions) is \conptime-complete with respect to the size of respective bounds of the \abstr. For first-order \abstr{s}, these problems are only partially decidable, but not  decidable.

Let us now turn to the entailment in the context of tightest \abstr{s}. If a source theory, a bridging theory, and an abstract vocabulary are given as input, checking the entailment of necessary or sufficient conditions is \conptime-complete with respect to the size of the computed tightest \abstr. Note, however, that the size of this tightest \abstr may be exponential in the size of the input theories. This complexity result holds both when using definitions~\eqref{eq:exists}, \eqref{eq:forall} as well as when applying Ackermann’s lemma for second-order quantifier elimination.

For the first-order case, the problem remains partially decidable  but not decidable whenever the relevant second-order quantifiers can be eliminated. If second-order quantifiers cannot be eliminated, the problem becomes  totally undecidable.\footnote{By \emph{totally undecidable} we mean here that the problem is not located at any finite level of the arithmetical hierarchy of Kleene and Mostowski (for a definition of the hierarchy see, e.g.,~\cite[Chapter 5]{hinman}).}

The problem of reasoning by querying belief bases can be formulated as follows~\cite{AHV}.
Given a fixed belief base consisting of a finite set of facts (either propositional variables or ground atomic formulas, i.e. predicates applied to constants), and a formula possibly containing free variables,
the task is to compute all substitutions of the free variables by constants occurring in the belief base such that the instantiated formula is satisfied (i.e. becomes true) in the belief base.\footnote{In the purely propositional case there are no variables to substitute, so the result of the query is simply either \true or \false.}

In the propositional case, this query-answering problem can be solved in linear time with respect to the size of the formula. In the first-order case, provided that no second-order quantifiers are present, query answering can still be performed efficiently: it is in \ptime and, in fact, in \logspace. This stands in sharp contrast to entailment, which for first-order formulas is only partially decidable.

When second-order quantifiers are present, the complexity rises:
\begin{itemize}
\item if existential second-order quantifiers ($\exists$) occur, as when applying Lemma~\ref{lemma:wsc-snc-so} to compute $snc()$ (and their elimination fails), the query-answering problem becomes \nptime-complete;\vspace*{-0.5em}
\item 
if universal second-order quantifiers ($\forall$) occur, as when applying Lemma~\ref{lemma:wsc-snc-so} to compute $wsc()$ (and their elimination fails), the query-answering  problem becomes \conptime-complete. 
\end{itemize}

\subsection{Remarks}

In the case of entailment, the computational complexity is inherited directly from the underlying logic -- propositional, first-order, or second-order, used to express the knowledge base and the query. Consequently, propositional entailment is \conptime-complete, first-order entailment is partially decidable but not decidable in general, and entailment with second-order quantification can be undecidable or even outside the arithmetical hierarchy, depending on the fragment considered.

When abstraction is applied to queries expressed as formulas/theories, the practical complexity is often much lower than what worst-case results might suggest. In many cases, the computational cost becomes comparable to that of query answering in \sql-like relational databases, which are routinely used in large-scale industrial applications. This observation indicates that abstraction-based reasoning by querying can be computationally friendly enough for practical use, even over large belief bases.

A significant body of work has demonstrated the effectiveness of Ackermann-style second-order quantifier elimination techniques in this context. In particular, the works~\cite{Del-PintoS19,ZhaoSchmidt16b,ZhaoS17}, along with other papers by these authors, show that such methods provide powerful tools for computing forgetting (also known as uniform interpolation) in a range of description logics. Their experimental results further demonstrate that these approaches are not only theoretically sound but also highly efficient and thus suitable for real-world reasoning tasks, such as ontology refinement, modularization, and privacy-preserving knowledge sharing. For a broader discussion of related techniques see also~\cite{dsz-forgetting-aij,gss}.

When applications of Ackermann’s lemma fail, i.e., when second-order quantifier elimination cannot be achieved by Ackermann-style substitutions, the fixpoint lemma of~\cite{NS} can be employed as an alternative. This lemma was originally developed in the context of modal correspondence theory, where it was used to characterize modal formulas via least and greatest fixpoints. Its relevance here stems from the fact that it provides fixpoint characterizations for broad classes of formulas that lie beyond the scope of the original Ackermann’s lemma. This makes it a valuable tool for querying belief bases, particularly in situations where queries or abstractions require a fixpoint definition. Importantly, the problem of evaluating fixpoint queries is known to be in \ptime, which means that reasoning based on the lemma remains computationally tractable~\cite{AHV}.
Moreover, the fixpoint lemma has been successfully applied to the computation of weakest sufficient  and strongest necessary conditions~\cite{dlssnc}. Since these notions are central to our framework, the lemma is directly applicable to the problems we consider.

We do not reproduce the formal statement of the lemma here, as its use would require extending the language to fixpoint logic, which goes beyond the scope of logics considered in this work.

\section{Conclusions}\label{sec:concl}

In this work, we proposed a new perspective on abstraction through the notion of \abstr and the bridge-and-bound methodology, where abstraction is based on selecting an abstract vocabulary and providing a propositional or first-order theory that bridges the lower level with abstract concepts. The bounds are then defined with respect to the bridging theory and abstract vocabulary. In addition to standard reasoning tasks focused on deriving necessary conditions, our framework also incorporates sufficient conditions as first-class entities. We believe that this approach models the process of abstraction more directly and naturally than existing models. It reflects a human-like way of reasoning and understanding complex systems and environments.

We began with the most general view of \abstr as arbitrary approximations of a~source theory, represented as a pair of theories (lower and upper bounds). We then introduced and studied tightest and exact abstractions, providing formal characterizations for both. In particular, we showed how tightest abstractions can be computed using weakest sufficient  and strongest necessary conditions, and we discussed the computational complexity of the relevant reasoning tasks.

For the sake of clarity of exposition and to maintain a well-focused scope, the framework presented in this paper is based on classical propositional and first-order logic. However, our approach is open to generalization: in the most important cases, namely those involving tightest and exact \abstr{s}, it relies solely on weakest sufficient and strongest necessary conditions, which can be defined in most other logics as well. As such, it can naturally be extended to other logic-based formalisms, including description logics, Answer Set Programming, modal logics, probabilistic logics, and more, where suitable notions of $wsc()$ and $snc()$ can be defined. This opens the door to analyzing abstractions in non-classical logics and rule-based systems.

Several promising research directions follow from this work. 
One direction is to provide sufficient conditions for the exactness of \abstr, other than based on definitional theories considered in Theorem~\ref{thm:exact}. 
Another is to study Theorem~\ref{thm:exact} in the context of automated abstraction techniques, for instance by discovering recurring patterns. 
Finally, Theorem~\ref{thm:layered} could be explored in the context of explainable AI, where layered abstractions may offer a principled way to generate human-understandable explanations.

\section*{Acknowledgments}

This work was partially supported by the ELLIIT Network Organization for Information and Communication Technology, Sweden, during the author’s second affiliation with the Department of Computer and Information Science, Link\"{o}ping University, Sweden.

\vfil
\eject
%\bibliographystyle{abbrv}
%\bibliography{abstr}

\let\oldbibitem\bibitem
\def\bibitem{\parskip=0pt\oldbibitem}

\end{document}